# Unified Mechanics on Thermodynamics, Classical Mechanics, and Quantum Mechanics

Henmei Ni *

School of Chemistry and Chemical Engineering, Southeast University, Southeast University Rd. 2#, Jiangning, Nanjing 211189, China. * henmei_ni@seu.edu.cn

## Abstract

A new thermodynamic framework shapes modern physics: electrons condense to form hot protons and cool protons (neutrons), then construct the nucleus through a thermal-difference force. An atom is an adiabatic electron-gas system. Heat is the sum of variations in kinetic energy, system potential, and environmental potential within a system. Force is the heat gradient over distance in the direction of transfer. Two kinds of forces exist between two particles, resulting from the relative thermal difference and the relative difference in number-volume density. Mutual repulsion and attraction simultaneously exist. Interaction doesn't need a medium. As a result, a series of concise equations is derived that includes only spatial parameters, mass, and temperature. Electricity and magnetism are nonessential concepts. 1) The equilibrium distance of molecular interaction: $L_e = \sqrt[3]{\frac{3\pi^{\alpha-1} m_A g}{4 N_A k T}}$; 2) the atom structure is built based on the radii of atoms and the temperature of electrons. An atom has only one isothermal orbital and several non-isothermal orbitals constituting a tetrahedron. Each orbital accommodates 2 electrons with opposite spins; 3) the 1st ionization energies (IE) are close to the kinetic energies of surface electrons, $\frac{3}{2}kT_e$; 4) the strengths of chemical bonds and H-bonds are calculated, e.g., the strength of O-O: $\Delta E = 2 \cdot \frac{3k}{2}(\frac{2}{\sqrt{3}} - 1)T_0$; 5) Light is heat transferred from a hot electron to a cold electron. The rate of hot electrons determines the velocity, but the hot molecule's rotation rate determines the frequency. For example, a $H_2$ on Earth transfers heat to a cold particle with velocity, $v_{exo} = 9.59 \times 10^7 sin(\pi f_e t)\cos(\pi f_e t)\cos(2\pi f_p t)$. Hence, the energy is transferred by piece and piece. This is the so-called wave-particle duality.

## Introduction

Physical principles should be founded on the ultimate and axiomatic concepts or hypotheses, i.e., mass ($m$), spatial dimension (e.g., point), and time ($t$), for falsification. Otherwise, the principles will fall into specious metaphysics. For example, it is scientific that a straight line of length $L$ is composed of an infinite number of points. The concept of length is well-defined, with the sub-concept of a point. However, it is metaphysical when we operate on a point, for instance, expanding a point or curving time, because there are no sub-concepts under a point or time. Another example is energy ($E$), a commonly used concept that is often confusing. Energy is one of the properties of a moving object with mass. In other words, mass is the carrier of energy. So far, only one is well-defined, i.e., the kinetic energy ($E_k$) of $\frac{1}{2}mv^2$, an eigen property of mass, where $v$ is the relative velocity of mass. This is a sub-concept of energy, on the hypothesis that the world consists of particles with mass. Potential energy ($E_p$), an eigen property of space, stemming from force, is only half-defined because force is not ultimately defined in terms of kinetic energy. Heat ($Q$), a kind of energy flowing between two systems, is completely undefined. Therefore, as a collective concept of energy is applied to describe a system, the physical connotation is specious. For example, Einstein's famous mass-energy equation, $E=mc^2$, where $c$ is the light's velocity, and the photoelectric equation, $E=hf$, where $h$ is Planck's constant and $f$ is the light's frequency, seem to describe heat substantially. However, they lack a relationship with variations in the kinetic and potential energies of particles in a system; in addition, their counterparts don't appear in the system that gains those energies; therefore, they are specious. Concretely, the mass-energy equation states that a system loses $m$ to deliver $mc^2$. However, the relation between $mc^2$ and the variation in the kinetic and potential energies of particles is not given, which is needed to determine the net energy the system releases. For example, if the law of energy conservation is applicable, $E_{exo} = mc^2 + \Delta E_k + \Delta E_p$. In addition, the counterpart of $mc^2$, i.e., the item that energy converts into mass, cannot be found in the energy-gain system. Similarly, the photoelectric equation must relate to variations in the kinetic and potential energies of electrons.

Kinetic energy is relative because velocity is measured relative to a selected, steady reference. The choice of reference is key to measurement. However, it is not sufficient. As shown in Fig. 1, to measure the velocity, the precondition is to distinguish between the environment and the systems. The environment refers to an infinite space and remains steady relative to the systems within it. Systems are shaped by their environments and constrained in a limited space. Moreover, the position of the light source (the torch in Fig. 1) matters, as does the observer's position, on the hypothesis that light or wireless is the only means of measurement. Fig. 1I shows a typical selection of the steady reference in physics: the torch and coordinate origin (observer) are fixed, and in most cases, their positions overlap. In this case, the relative velocities of other systems in the environment are continuously measurable, no matter where the observer stands. For example, when measuring planetary orbits in the solar system, the light source (torch) and the coordinate origin are set at the Sun. Alternatively, the observed systems have their own torches to replace torch O. In contrast, as shown in Fig. 1II, solely moving the coordinate origin to system A, i.e., the unfixed torch and the coordinate origin, returns to the geocentric model in the mid-century: the continuity of measurement is broken because the reflection of the planet instantly and randomly changes. For example, we observe the motion of satellites around the Earth depending on sunlight reflected off solar cells. The following result is not ridiculous: the satellites flash and jump because they are observed only when the tiny reflection plane faces the observer. Namely, if the satellites are indistinguishable, they appear and disappear at random in very short time intervals. Accordingly, the probability of an emerging position best describes the observation results. However, the defect is that the spatial constraints on motion are hard to determine. As shown in Fig 1II, the probability depends on the positions of the observer and the torch. Therefore, the observation results may not reflect the true motion. Unfortunately, a similar situation occurs in quantum theory, where results depend on electrons' responses to photons outside an atom. Quantum theory has led to major advances in engineering. However, like geocentric theory, the complicated mathematical computations imply that these achievements likely result from rhythmic electron motion rather than uncertain probability. One piece of evidence is the uncertainty principle. Probability means uncertainty due to a lack of spatial constraints. However, if we attribute this uncertainty to photons interacting with tiny electrons, we need a definition of light. Undoubtedly, light is heat, an energy transferring between two objects, which is so far undefined. Hence, it is worth rechecking quantum theory. A basic indicator of success is to correlate an electron's kinetic energy, a property of mass, with the emission energy, if the emission is the behavior of electron motion.

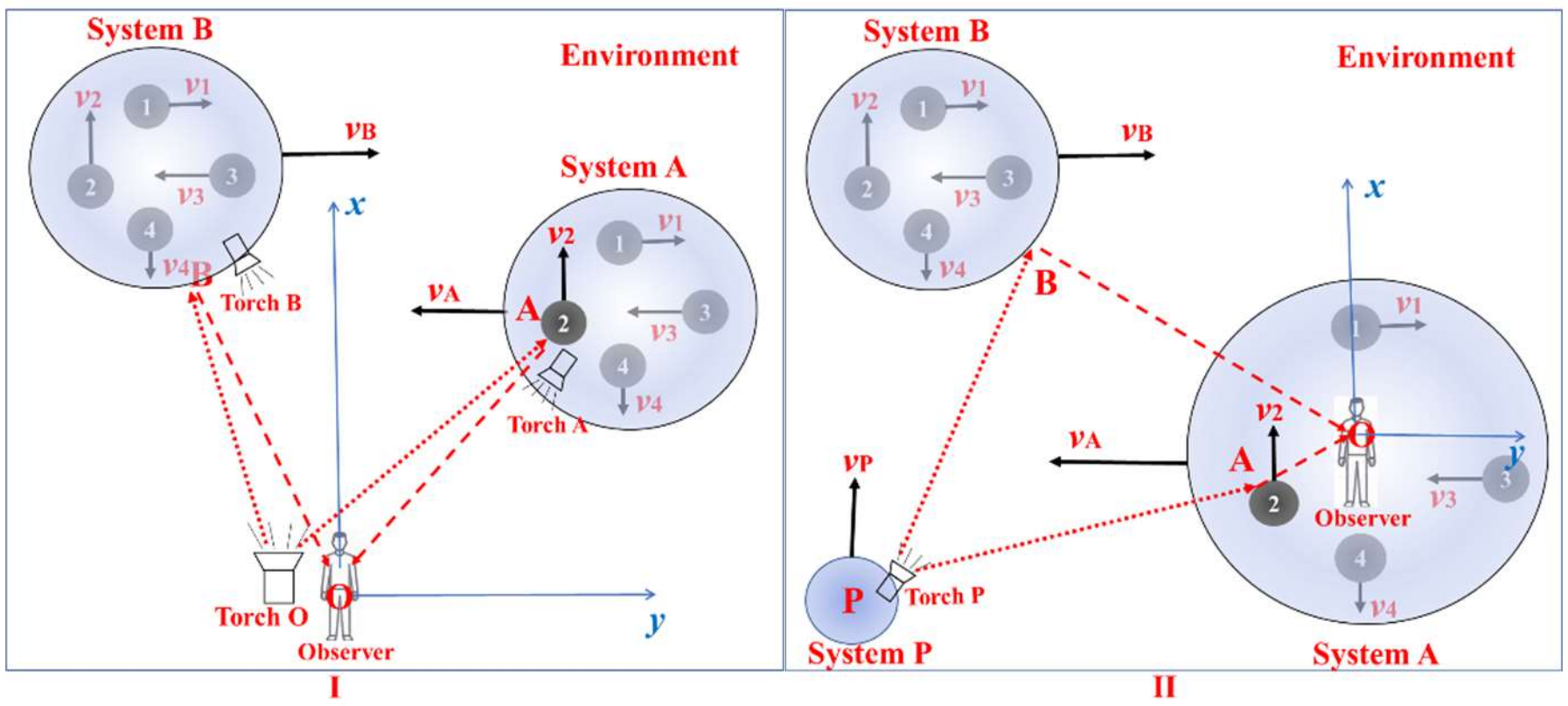

Figure 1. The measurements of velocity

Temperature is defined by the average kinetic energy of particles, which can also be extrapolated to a single particle as $k_BT$, where $k_B$ is the Boltzmann constant, and $T$ is the particle's temperature, derived from its velocity. Therefore, temperature is also relative. For example, the gas temperature in a container usually reflects the average kinetic energy of gas molecules relative to the hypothetical planes of the container walls, rather than the atoms in the walls. This matters because the temperature difference between the atoms in the container walls and the gas molecules determines heat transfer. Therefore, it should be arguable what the significance of the so-called absolute temperature of zero is. By definition, temperature may be zero because of the choice of coordinate origin, and the temperature difference may be zero too, indicating that the two objects are at rest relative to each other. Hence, the absolute zero temperature should imply an ultimate state, e.g., the origin of the cosmos, at which all the fundamental particles are at rest relative to each other. For example, assuming that electrons are the fundamental particles, with the name electron, which is neutral in electricity, in the initial cosmos, all the electrons are at rest. Due to whatever disturbance, the temperature rises, triggering the formation of protons and neutrons.

On the other hand, force is a concept commonly used in daily life and academics. However, it is not defined in Newtonian classical mechanics, electrodynamics, or quantum mechanics. Newton stated that when an object changes its state of motion, a force is exerted on it. However, Newton did not explain the cause of force creation or the property of a force. Without this concept, we cannot define work and potential energy. Undoubtedly, classical mechanics has achieved fruitful gains that have advanced civilization. However, it is a collection of observational results, such as universal gravitation, the law of inertia, and angular momentum conservation, rather than a theoretical system.

Heat is a kind of energy transferred between two systems. However, it is not defined. This is a fatal defect in classical thermodynamics because thermodynamics is substantially the science of heat. Classical thermodynamics completely separates work and heat in its definition of internal energy and attempts to determine the quantity of heat by extracting measurable work from internal energy.

However, since the relationship between state parameters, namely pressure, temperature, and density, and heat is not established, it is questioned whether heat has anything to do with the variations of state parameters. Entropy is suggested to correlate heat and temperature. However, entropy cannot yet be computed. Statistical thermodynamics does much work to compute the translational, rotational, and vibrational energy of molecules or particles, but cannot establish a microscopic relation with the work a system does. This is because neither heat nor force is defined. In addition, classical thermodynamics does not include gravitational potential energy in its equations. However, heat transfers through three modes: thermal conduction, irradiation, and thermal convection. Gravitation causes thermal convection. Therefore, classical thermodynamics, including statistical thermodynamics, has nothing to do with heat. This may be because the drive to extract more useful work from thermal machines spurred research in thermodynamics, but it artificially splits work and heat, which a system exports as the same thing. As a result, heat loses its definition.

In summary, although human beings have gained huge achievements from modern science, scientific theories are incoherent. Therefore, a unified theory is necessary.

In the previous papers [1]-[3], inspired by the equations' induction for gas solubility, gas-liquid interfacial tension, and interfacial pressure, a heat-transfer path was suggested, in which heat transfers from a gas system to the surroundings via gas molecules that undergo inelastic collisions with the container walls and then liquefy. This suggestion led to a new, concise gas-state equation for the gas-liquid saturation system [2], eliminating the distinction between ideal and real gases. Moreover, the authors established a new model for the specific heat capacities of gas at saturation [3]. These computational results are remarkably consistent with the standard data issued by the National Institute of Standards and Technology, USA (NIST). Therefore, this paper further applies the heat-transfer model to classical and quantum mechanics.

## Results and Discussion

## Definition of Heat

A hypothesis is the foundation of any theory. As shown in Fig. 2, the initial cosmos is full of fundamental particles with mass, called electrons, which are electrically neutral. Because of some disturbance, the cosmos (environment) shapes the systems step by step (Fig. 2 I→III), i.e., protons, neutrons, atoms, etc.

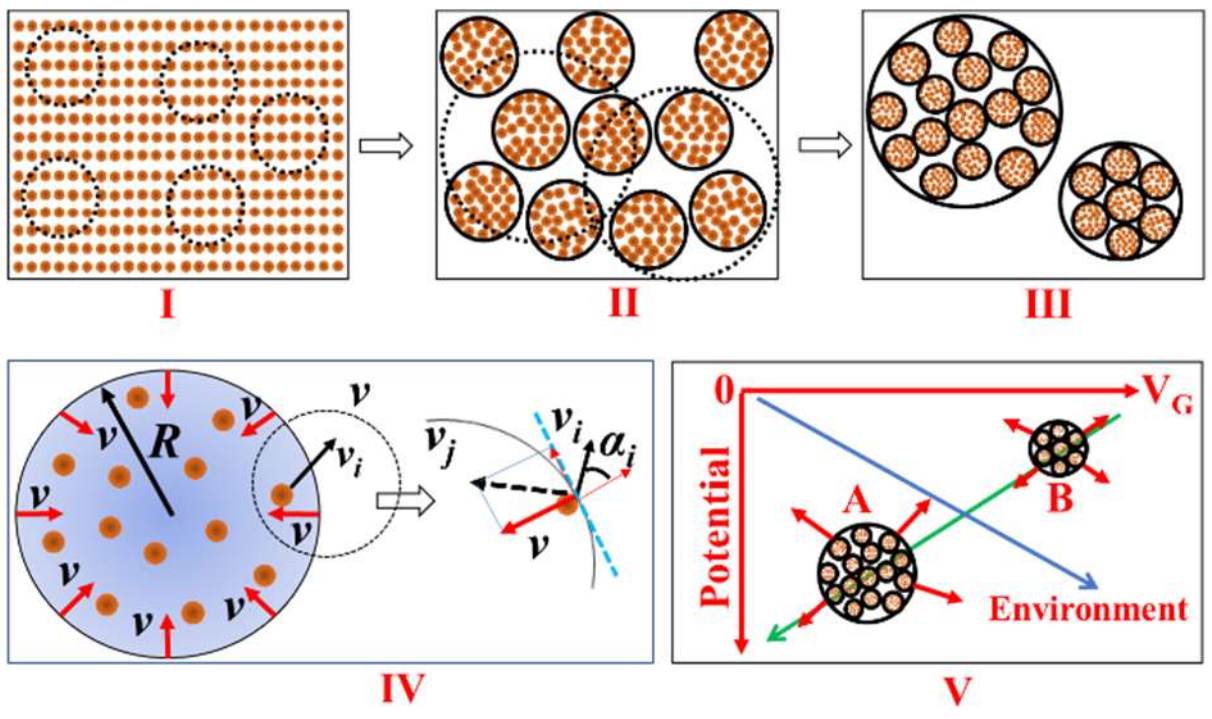

Figure 2, Scheme of systems created by the cosmos

Fig. 2 IV shows an example of a process in which a system is shaped under adiabatic conditions. When a system is compressed with a constant velocity $v$, the kinetic energy of particles increases.

$$v_j^2 = (v + v_i cos\alpha_i)^2 + v_i^2 sin^2\alpha_i = v^2 + v_i^2 + 2vv_i cos\alpha_i \qquad (1)$$

$$\sum_{j=1}^{n_i} v_j^2 = \sum_{i=1}^{n_i}(v^2 + v_i^2) + 2v\sum_{i=1}^{n_i} v_i cos\alpha_i \qquad (2)$$

Here, $v_i$ is the velocity of a particle or a sub-system relative to the sphere center, while $n_i$ is the number of particles colliding with the boundary at time $t_i$; $v_j$ is the velocity of a particle after collision; and $\alpha_i$ is the angle between the velocity and the normal line (radius). When $n_i$ is large enough and the velocity directions are random, Eq 2 can be approximated by the average velocity: $0 \le \alpha_i \le \frac{\pi}{2}$,

$$n_i\overline{v_j}^2 = n_i\overline{v_i}^2 + n_i\left(v^2 + \frac{4}{\pi}v\overline{v_i}\right) \qquad (3)$$

$$\overline{v_j}^2 - \overline{v_j}^2 = v^2 + \frac{4}{\pi}v\overline{v_i} \qquad (4)$$

If a particle has mass $m$, the system's kinetic energy increases over time.

$$dE_k = \frac{m(v^2 + \frac{4}{\pi}v\overline{v_i})}{2} \cdot dn_i \qquad (5)$$

If there are $N$ particles in a system, and each particle collides once with the boundary, $N$ particles have $\overline{v_j^2}$. Since only half of the particles near the boundary collide,

$$dn_i = -\frac{1}{2} \cdot \frac{3N}{R} dR \qquad (6)$$

$$\int_i^j dE_k = -\frac{3Nm(v^2 + \frac{4}{\pi}v\overline{v_i})}{4}\int_{R_i}^{R_j}\frac{dR}{R} = \frac{3Nm(v^2 + \frac{4}{\pi}v\overline{v_i})}{4} ln\frac{R_i}{R_j} = \frac{3Nm(v^2 + \frac{4}{\pi}v\overline{v_i})}{4} ln\frac{V_{G,i}}{V_{G,j}} \qquad (7)$$

Here, $V_G$ indicates the molar volume. Therefore,

$$\Delta E_k = \frac{3Nm}{4}(v^2 + \frac{4}{\pi}v\overline{v_i}) \cdot ln\frac{V_{G,i}}{V_{G,j}} \qquad (8)$$

The time interval is $\Delta t = \frac{R_i - R_j}{v}$.

Assuming that the compression rate simultaneously fits the variation of the average velocity, for instance, $v = \overline{v_i}$, then,

$$\Delta E_k = (1 + \frac{4}{\pi})\frac{3Nm\overline{v_i}^2}{4} \cdot ln\frac{V_{G,i}}{V_{G,j}} \qquad (9)$$

Eq 9 shows the case of adiabatic compression. If the system exchanges kinetic energy with the environment, e.g., $v_i cos\alpha_i = 0$ (Fig 2 IV),

$$v_j^2 = v^2 + v_i^2 sin^2\alpha_i \qquad (10)$$

$$\overline{v_j}^2 = v^2 + \frac{\overline{v_i}^2}{2} = \frac{3}{2}\overline{v_i}^2 \qquad (11)$$

Eq 9 deforms to be, using the definition of temperature in classical thermodynamics, $Nm\overline{v_i}^2 = 3RT_i$,

$$\Delta E_k = \frac{3Nm\overline{v_i}^2}{8} \cdot ln\frac{V_{G,i}}{V_{G,j}} = \frac{9RT_i}{8} \cdot ln\frac{V_{G,i}}{V_{G,j}} \qquad (12)$$

Here, $N$ is one mole. Eq. 12 gives the increase in kinetic energy as temperature rises from $T_i$ to $T_j$.

It should be noted that, in the above inductions, the collisions of particles inside the system are regarded as the only path for the kinetic energy exchange among the particles.

Now, let's focus on kinetic energy exchange with the environment. As shown in Fig 2. IV, as $v = 0$, $v_i cos\alpha_i$ is the unique item governing the exchange of kinetic energy. $v_i sin\alpha_i$ leads the particle to move on the boundary surface. Many particles moving on the surface form the liquid phase. Namely, the particles collide with each other on the surface, but do not enter the gas phase. Conversely, when the environment transfers $v_i cos\alpha_i$ to a particle in the liquid phase, the particle moves back to the gas phase. In this way, the energy-exchange cycle forms.

However, liquefaction decreases the number of particles in the gas phase. As a result, $V_{G,j}$ increases to $V_{G,i}$, but $T_j$ descends to $T_i$. Therefore, to keep the temperature, the environment should boost liquid particles with a velocity of $v_i cos\alpha_i$.

$$\Delta E_k = \frac{9RT_i}{8} ln\frac{V_G}{V_L} \approx RTln\frac{V_G}{V_L} \qquad (13)$$

Accordingly, at thermal equilibrium, if a gas system loses $\frac{3}{2}RT$, it must gain $RTln\frac{V_G}{V_L}$. Conversely, if a system obtains $\frac{3}{2}RT$, it must exhaust $RTln\frac{V_L}{V_G}$ to keep the temperature.

Since the kinetic energy is originally zero (T=0) in the cosmos, when the spatial constraint power creates kinetic energy, a trend emerges for all particles. If the shaping power disappears, as shown in Fig 2 V, all particles spontaneously recover their original state. This may be the origin of the 2nd law in classical thermodynamics. Hence, the counterpart of kinetic energy is needed to present the trend. This is the potential energy. $RTln\frac{V_G}{V_L}$ is defined as the potential energy that the environment stores in a gas system.

On Earth, gravitational potential is well known, but no one has yet given a mathematical expression for it in a gas system. In our previous paper [1]-[3], we proposed an expression and successfully applied it to the equation of state of a real gas. It is the environmental potential loading on any system, which relates to the relative height and number of particles. It rises as the volume increases. Therefore, we use it to define heat.

In summary, adopting the standard in classical thermodynamics that the exothermic is positive, and the endothermal is negative, at heat equilibrium, the exothermic heat of a gas system is defined as,

$$Q_{exo} = \frac{3}{2}RT - RTln\frac{V_G}{V_L} - \frac{\pi^\alpha mg}{V_G} \qquad (14)$$

where $R$ is the gas constant (8.314 $J{\cdot}K^{-1}{\cdot}mol^{-1}$), $T$ is the temperature (K), and $V_L$ and $V_G$ are the molar volumes of liquid and gas (L/mol), respectively. $\pi$ is the circumference constant (3.14), $\alpha$ is an arbitrary factor, which has been given in Ref. [2], $m$ is the molecule's molar mass, and $g$ is the gravitational constant (9.8 $m/s^2$). It should be noted that the heat emitted by chemical reactions and irradiation is not included in Eq 14. Chemical reactions and irradiation are behaviors of the subsystems, i.e., the atoms and electrons. Although subsystem behavior is closely related to system behavior, we first consider system behavior for succinctness.

Inversely, the endothermic equation of a gas system is, by simplification, $\beta = \frac{\pi^\alpha mg}{V_G RT}$,

$$Q_{endo} = -\frac{3}{2}RT - RTln\frac{V_G}{V_L} + \beta RT \qquad (15)$$

It should be noted that, in Eq 15, $-RTln\frac{V_G}{V_L}$ initially is $RTln\frac{V_L}{V_G}$, indicating the exhausted potential energy for keeping the temperature. Meanwhile, a problem arises: $Q_{endo} + Q_{exo} \neq 0$.

$$Q_{exo} + Q_{endo} = -2RTln\frac{V_G}{V_L} \quad (16)$$

We think that Eq. 16 reflects the environment's shaping power. Without this power, a closed or isolated system cannot form.

On the other hand, Eq 16 resembles the Gibbs energy. Gibbs is correct that at equilibrium the free energy is minimal. However, Gibbs' theory cannot determine that minimum value. Therefore, Eq. 16 shows that classical thermodynamics complicates things. Concepts such as internal energy, enthalpy, entropy, and free energy are not necessary.

Eq. 14 and Eq. 15 are heats at thermal equilibrium. Here, we discuss a cylinder with a piston under adiabatic conditions. When an ideal elastic collision happens between a particle and the piston, the particle doesn't lose any kinetic energy. Moreover, potential energy is the counterpart of kinetic energy. It doesn't work if kinetic energy doesn't change. Therefore, the piston doesn't gain energy to move. Only when the piston gains the kinetic energy of the gas molecules, i.e., $v_i cos\alpha_i$, does it move. Under adiabatic conditions, the environment cannot supply energy to recover the gas temperature in the cylinder. Hence, the gas system loses kinetic energy, while the environment gains it and its counterpart, potential energy. This potential energy is called work.

## Definition of Force

As we point out, classical mechanics does not define force. Newton described the effect after the interaction of two objects.

$$\vec{F} = \frac{d\vec{p}}{dt} = \frac{md\vec{v}}{dt} \quad (17)$$

Here, $p$ denotes momentum. It states that if an object's velocity varies with time, a force is exerted on it. Eq. 17 cannot predict what happened before the force was created. Moreover, it doesn't determine the property of force. However, it implies that the force transfers kinetic energy to the object. Therefore, we think that force is a property of heat. Since force has a point-type property, force is defined as the linear gradient of heat along the direction of heat.

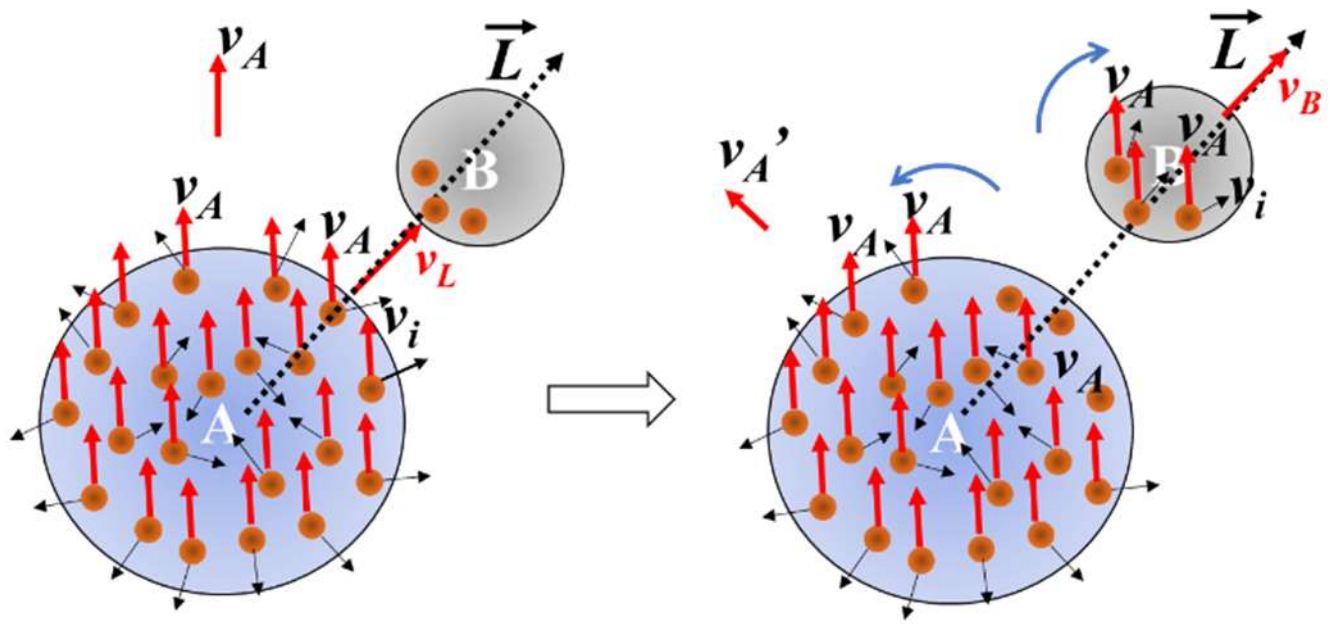

Figure 3, Scheme of collision

Fig 3 shows a scheme of collision between two objects. When A and B contact, in addition to the system A and B exchanging kinetic energy (i.e., a portion of $\frac{1}{2}m(v_A^2 + v_i^2)$), the subsystems in both A and B exchange kinetic energy. The $v_B$ direction is determined by the site that A and B touch, i.e., $v_L$. Hence, it is predictable by the spatial positions of A and B, as well as the relative directions of motion. Eq 17 can describe $v_B$, namely that the direction of $F$ is consistent with that of $v_B$. However, it cannot describe A's motion because the direction of $v_A$ is not consistent with that of $F$. Consequently, the direction of heat transfer, i.e., $\vec{L}$, is the best choice to determine the direction of force. As shown in Fig. 3, collision-induced motion is complicated. It is not purely straight motion, as a point particle described by Newtonian laws would be. Spinning or rotating is possible. Moreover, in classical mechanics, we apply conservation of kinetic energy and conservation of momentum in collisions. However, this is questioned because heat is not negligible, which leads to a temperature rise. Therefore, a new concept of force is essential.

At thermal equilibrium, the force of exothermic heat is:

$$\overrightarrow{F_{exo}} = \frac{dQ_{exo}}{d\vec{L}} = R\left(\frac{3}{2} - ln\frac{V_G}{V_L}\right) \cdot \frac{dT}{dL}\vec{L} - (1-\beta)RT \cdot \frac{dlnV_G}{dL}\vec{L} - RT\frac{dlnV_L}{dr}\vec{r} \qquad (18)$$

$\vec{L}$ and $\vec{r}$ are the unit vectors. *L>>r*, where *L* and *r* are the distances that a gas particle is from the center of the liquid phase and the radius of the liquid phase, respectively. *dL=-dr* because the radius increases.

Similarly, the endothermal force is the linear differentiation of Eq. 15,

$$\overrightarrow{F_{endo}} = -R\left(ln\frac{V_G}{V_L} + \frac{3}{2}\right) \cdot \frac{dT}{dL}\vec{L} - (1+\beta)RT \cdot \frac{dlnV_G}{dL}\vec{L} - RT\frac{dlnV_L}{dr}\vec{r} \qquad (19)$$

Furthermore, the pressure of a gas system is,

$$P_{exo} = \frac{dQ_{exo}}{dV_G} = R\left(\frac{3}{2} - ln\frac{V_G}{V_L}\right) \cdot \frac{dT}{dV_G} - (1-\beta) \cdot \frac{RT}{V_G} - RT\frac{dlnV_L}{dV_G} \qquad (20)$$

$$P_{endo} = \frac{dQ_{endo}}{dV_G} = -R\left(ln\frac{V_G}{V_L} + \frac{3}{2}\right) \cdot \frac{dT}{dV_G} - (1+\beta) \cdot \frac{RT}{V_G} - RT\frac{dlnV_L}{dV_G} \qquad (21)$$

Under conventional conditions, the minor $V_L$ variation leads to a negligible $RT\frac{dlnV_L}{dV_G}$. Therefore, if neglecting the gravitation and $\frac{dT}{dV_G} = 0$, i.e., the ideal condition in classical thermodynamics: $P = \frac{RT}{V_G}$.

In addition,

$$P_{exo} + P_{endo} = -2Rln\frac{V_G}{V_L} \cdot \frac{dT}{dV_G} - \frac{2RT}{V_G} \qquad (22)$$

It indicates that the external and internal pressures of a gas system differ. Since the value is negative, the environment exerts more pressure on the gas system.

A special case is a gas in a cylinder with a movable piston (neglecting its weight). The gas system transfers its potential energy $RTln\frac{V_G}{V_L}$ to the environment as work, but without the reward from the environment. It indicates that the useful work is a part of heat. Hence, the definition of internal energy is unreasonable in classical thermodynamics. The exothermic equation is, but the endothermal equation doesn't change,

$$Q_{exo} = \left(\frac{3}{2} + ln\frac{V_G}{V_L}\right)RT - \beta RT \qquad (23)$$

Hence, $Q_{exo} + Q_{endo} = 0$. However, as the system loses kinetic and potential energies, the temperature decreases, whereas the environment gains work and kinetic energy; therefore, Eq 23 is an instantaneous equation. Meanwhile, it shows the property of force: only the system's potential force does work, while the other forces have other functions. Certainly, the endothermal equation doesn't exist because the environment provides no rewards. Therefore, the gas system is expected to degrade until it returns to the environment. Namely, the environment retires its shaping power.

Another special case is when the cylinder is adiabatic, and the environment is a vacuum. Eq. 22 = 0 may describe the instantaneous state of the gas, where the liquid term is neglected.

$$dlnV_G = -ln\frac{V_G}{V_L} \cdot dlnT \quad (24)$$

It interprets the Joule-Thomson effect.

In summary, item $ln\frac{V_G}{V_L}$ is the largest among the above equations, while the kinetic energy item is the second-largest. This shows that kinetic energy variation is only an initiating factor that drives the expansion or contraction of the space. The system potential contributes most of the heat and force. It is something like a traffic accident occurring on the highway. As soon as one or two cars break down, all the following cars have to stop. Hence, the following cars release most of the heat.

## Extrapolating Thermal Forces to Classical Mechanics

Thermodynamic equations reflect the collective behaviors of particles in a system. Hereafter, we try to extrapolate the force equations to treat the interaction between two objects. Since the force equations contain many terms, we analyze them one by one.

### 1) Force derived from kinetic energy

This item is derived from the collision.

$$\overrightarrow{F_{exo}} = \frac{3}{2}k \cdot \frac{dT}{dL}\vec{L} \quad (25)$$

Here, $k$ is the Boltzmann constant for a particle. $\frac{3}{2}k\Delta T_{AB} = \frac{1}{2}m_A v_{L,A}^2 - \frac{1}{2}m_B v_{L,B}^2$, where $m_A$ and $m_B$ are the mass; $v_{L,A}$ and $v_{L,B}$ (regarded as zero) are the velocities of A and B along $\vec{L}$, respectively. $\Delta T_{AB}$ is the A-B temperature difference.

$$F_{exo} = \frac{1}{2}\frac{d(m_A v_{L,A}{}^2)}{dL} = m_A a_{L,A} \quad (26)$$

As shown in Fig 3, since A actively collides with B, $a_{L,A}$ for A is a negative acceleration.

$$F_{endo} = -\frac{1}{2}\frac{d(m_A v_{L,A}{}^2)}{dL} = -m_A a_{L,A} \quad (27)$$

The equation is consistent with Newton's 2nd law in form, but differs in significance. Eq 26 gives a negative value, indicating a force in the opposite direction to $\vec{L}$ exerts on A, while Eq 27 shows that a force directed along $\vec{L}$ exerts on B. The property of interaction is repulsive.

Newton did not mention this process, but suddenly postulated the following relationship,

$$F_{L,B} = m_B a_{L,B} = -F_{exo} = -m_A a_{L,A} \quad (28)$$

Eq 28 bridges the gap between Newton's 2nd law and present theory in accordance with momentum conservation or kinetic energy conservation for two dimensionless objects.

Note that force can only be exerted on an object with mass; thus, artificially choosing a $\vec{L}$

directing vacuum is insignificant.

Finally, Newton's 3rd law states that action and reaction occur simultaneously in opposite directions with equal strength. However, from the standpoint of heat transfer, both Newton's 2nd and 3rd laws are suitable only for the collision between two dimensionless objects, not for the collision of real objects.

## 2) Temperature difference force

The second item, which is the largest, results from the system potential.

$$F_{exo} = -\frac{2}{3} ln\frac{V_G}{V_L} \cdot m_A a_{L,A} \tag{29}$$

Molar volume is a collective item. It is arguable whether it is appropriate to apply to two independent particles. Kinetic energy is predominantly transferred through collisions between independent particles. This item reflects a collective behavior of many particles in the temperature gradient. Reminiscent of a traffic accident on a highway, this item should be caused by the balance of the subject forces, rather than a collision. Anyway, considering the A→B system where B's temperature is zero, since there are only two particles, $V_G = \frac{4}{3} N_A \pi L^3$ ($N_A$ is Avogadro number) is designated to A, while $V_L$ is assigned to B.

$$F_{exo} = -2(lnL - lnr_B) \cdot m_A a_{L,A} \tag{30}$$

$r_B$ is the radius of B. According to Eq 28,

$$F_{L,B} = -F_{exo} = 2(lnL - lnr_B) \cdot m_A a_{L,A} = m_B a_{L,B} \tag{31}$$

Since $a_{L,A} < 0$, in addition to $L >> r_B$, $F_{exo} > 0$ and $F_{L,B} < 0$. It indicates that A and B are mutually attractive.

At thermal nonequilibrium, a system does work on the environment. Using Eq 23 for differentiation, a result similar to the force of kinetic energy is obtained: $F_{exo} < 0$ and $F_{L,B} > 0$, namely that A and B are mutually repulsive.

However, whether in equilibrium or non-equilibrium, the force has a special property: it becomes stronger as the distance increases. Since this force comes from the system potential, it recalls the situation shown in Fig. 2V. During the process of shaping a system, the spontaneous trend of subsystems is to return to the environment; thus, the environment naturally exerts a force on each subsystem. Moreover, remember that this comes from the equations at thermal equilibrium, where the potential is supplied by their opposites to maintain a constant temperature; thus, the environment intervenes in the interaction. Therefore, larger L means a larger volume and thus a higher potential; hence, the force is stronger. A strong force compresses a particle at dilute concentrations to gain kinetic energy, which averages the temperatures over a short time.

## 3) Number-Volume-density force.

$$\overrightarrow{F_{exo}} = -(1-\beta)RT \cdot \frac{dlnV_G}{dL}\vec{L} \tag{32}$$

It is the force created by the molar volume variations under isothermal conditions. In the above sections, $ln\frac{V_G}{V_L}$ is the most oversized item under conditions far from the critical temperature or pressure [29], [31]. Hence, the temperature-difference force is far more potent than the number-volume-density force, unless, near critical conditions, the number-volume-density force dominates.

A's $V_G$ decreases as $L$ decreases, $\frac{dV_G}{dL} < 0$. Here, it should be noted that the unit of $L$ is *dm* because the unit of $V_G$ is L/mol. Hence, at thermal equilibrium,

$$F_{exo} = \frac{3kT}{L} - \frac{9\pi^{\alpha-1} m_A g}{4N_A L^4} \tag{33}$$

As discussed above, the positive $\frac{3kT}{L}$ in Eq 33 indicates that A and B are mutually attractive. The negative $\frac{9\pi^{\alpha-1} m_A g}{4N_A L^4}$ shows the repulsive property. It is derived from the gravitational term, which is inversely proportional to the 4th power of distance. This means the environment potential creates a mutually repulsive force between two subsystems, whereas the system potential creates a mutually attractive force. On Earth, the gas system potential comes from Earth (environment); however, Earth's gravitational potential comes from the solar system, i.e., the environment at the next higher level relative to Earth. Hence, by analogy, the cosmos and galaxy may shape their subsystems with different potentials. On the other hand, number-volume-density, rather than mass-volume-density, determines the potential. It provides a new theoretical framework for investigating cosmic evolution.

Since $V_L$ normally varies little, the last item, i.e., $-kT\frac{dlnV_L}{dL}$, is negligible. In addition, A and B are exchangeable because of the isotherm.

Thermal-difference force has been applied in many fields. An example is the Wilson cloud chamber, used to detect high-energy particles such as α-particles and neutrons, in which high-speed particles attract molecules and condense them. In addition, gas mechanics is a discipline that studies the force due to thermal differences.

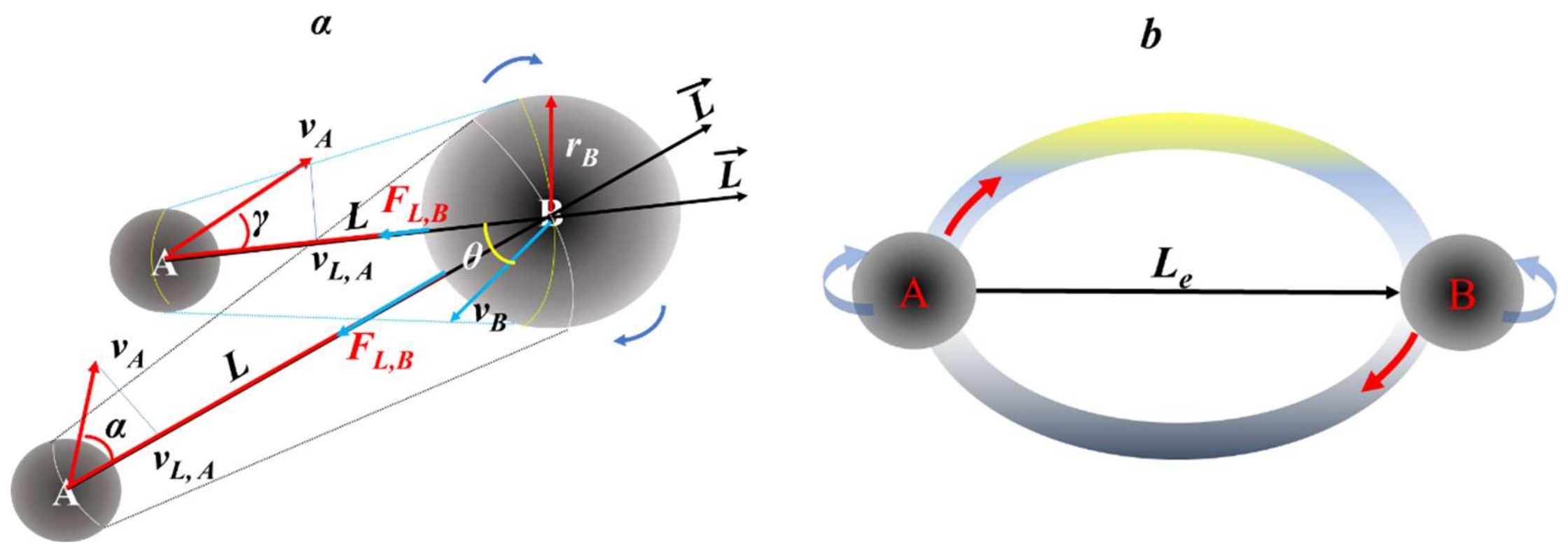

Figure 4, Scheme of A-B interactions

We can now summarize sections (1)- (3). As shown in Fig. 4a, A approaches B, which remains still. The force from the temperature difference is strong at a distance. It drives A and B to approach each other. The speed is determined by the angular rate shown in Eq. 34. As a result, the temperature difference decreases.

$$F_{L,B} = -2km_A ln\frac{L}{r_B}\cdot\frac{v_A dcos\alpha}{v_A cos\alpha dt} = 2km_A tan\alpha \cdot ln\frac{L}{r_B}\cdot\frac{d\alpha}{dt} \tag{34}$$

Since the computation is very complicated, we describe the results qualitatively.

a) When $\gamma < \frac{\pi}{2}$, and $L - r_B < 0$, A collides with B. The temperature difference is $v_{L,A}cos\gamma + v_{L,B}cos\theta$, where $\theta$ is the angle between $v_B$ and $L$.

b) When $\gamma < \frac{\pi}{2}$, $L - r_B > 0$, and $\frac{m_A v_{L,A}^2 cos^2\gamma}{m_B v_{L,B}^2 cos^2\theta} = 1$, the isothermal equation Eq. 33 works. Since the repulsive force is proportional to the -4$^{th}$ power of the distance and independent of temperature, it works even under non-isothermal conditions, but is negligible at long distances. At an isotherm, it blocks the collision. Hence, a question arises: what is called the "touch" of two objects? Action at a distance has confused modern physics for hundreds of years because physicists believe matter must act as a medium for force transfer. We now understand that force never moves; kinetic and potential energy do. Since both kinetic and potential energy are shaped by the environment, the environmental power forces subsystems to incorporate, whereas the upper-level environment forces subsystems to disperse. Space is a platform and systems are marionettes operated by the environments. Therefore, systems never "touch" each other even at so-called collision.

If following up, the ultimate question is: where does environmental power come from? This is equivalent to asking: how is the cosmos born? So far, one point seems clear: at the beginning, the sum of kinetic and potential energy, as defined by the human being, should be zero. Of course, we don't support the hypothesis of dark matter and dark energy because the evidence supporting them is interpretable in the present theory.

c) Since A and B are not dimensionless points, and force has the point property, as shown in Fig. 4a, the interaction should be surface-to-surface. We may call it tide force. All the points on the face-to-face surfaces of A and B simultaneously interact, and more: not one-to-one points, but one-to-all points interact. This directly implies that the potential and its force depend on number-volume-density rather than mass-volume-density because the total surface area is proportional to the number of particles, a principle we know from classical thermodynamics. Particles within a system simply constrain the surface particles. Mass is certainly important to the temperature-difference force; however, the force must be mediated through the surface particles. Kinetic energy is a sum of collective behaviors of particles in a system; thus, it relates only to the total number of particles, regardless of the volume.

Modern physics typically uses concepts like the gravitational field, electric field, and magnetic field to describe the potential distribution. We think of fields not as objective entities, but as mathematical tools for convenient computation. Fields don't exist for a single particle, though they can predict the energy level at any position when a given particle transfers heat to another particle.

d) Since the force strength decreases and the direction varies as A approaches B, the spin likely occurs due to B's geometry and/or the mass distribution.

e) Provided that A and B are two congeneric particles, the final orbital shown in Fig. 4b can form. The two objects chase each other in a cycle while spinning in opposite directions. The equilibrium distance, $L_e$, can be computed. Eq. 33=0, thus,

$$L_e = \sqrt[3]{\frac{3\pi^{\alpha-1} m_A g}{4N_A kT}} \tag{35}$$

We compute the equilibrium distances of two molecules in some pure gases at 298 K and list the results in Tab 1.

Table 1: Calculated equilibrium distance of gas molecules at 298 K

| Gas | α | $m_A$ (g/mol) | $L_e$ (Å) | Ref. Value | Gas | α | $m_A$ (g/mol) | $L_e$ (Å) | Ref. Value |
|---|---|---|---|---|---|---|---|---|---|
| **He** | **-1.83** | **4** | **0.92** | | **$CH_4$** | **0.6** | **16** | **3.68** | **3.74[35],3.71[38]** |
| Ne | -1.64 | 20 | 1.69 | 2.80[34] | $CH_3CH_3$ | 0.82 | 30 | 4.94 | 3.51[35],4.89[36] |
| Ar | -0.69 | 40 | 3.05 | 3.40[35] | $C_3H_8$ | 0.88 | 44 | 5.74 | 5.56[36] |
| Kr | -0.87 | 121.8 | 4.13 | 3.63[34], [38] | $C_4H_{10}$ | 0.95 | 58 | 6.46 | |
| Xe | -0.74 | 131.2 | 4.45 | 3.95[34], [38] | Ethylene | 0.67 | 28 | 4.56 | 3.54[38] |
| $H_2$ | 0.54 | 2 | 1.80 | | Benzene | 0.96 | 78 | 7.16 | |
| $N_2$ | -1/3 | 28 | 3.11 | 3.31[35], 4.13[36] | Toluene | 1.0 | 92 | 7.68 | |
| $O_2$ | -0.45 | 32 | 3.11 | 3.20[37],3.95[38] | $NF_3$ | -1/3 | 74 | 4.30 | |
| $F_2$ | -0.78 | 40 | 2.95 | 2.83[35] | $CF_4$ | -0.45 | 92 | 4.42 | 5.21[36] |
| CO | -0.29 | 28 | 3.16 | 3.17[38] | $SF_6$ | -1/3 | 152 | 5.46 | 5.18[36] |
| $CO_2$ | 0.07 | 44 | 4.21 | 2.94[35],4.38[36] | $CH_3OH$ | 0.93 | 32 | 5.26 | |
| $SO_2$ | 0.25 | 64 | 5.11 | | $NH_3$ | 0.87 | 17 | 4.16 | 3.90[36] |
| $H_2O$ | 0.95 | 18 | 4.38 | 3.57[36] | $H_2S$ | 0.49 | 34 | 4.54 | |

Note: $\alpha$ values have been given in Reference [29]

No standard data exist in the literature, though the van der Waals radii of elements were calculated from equilibrium distances [33]. However, the data in Tab. 1 are reasonable and consistent with the empirical use of equilibrium distances in molecular dynamics (MD) simulations [34]-[38].

Compared with the Lennard-Jones potential (Eq. 1) [11]-[14], the derivation of Eq. 35 is straightforward and concise without adopting any assumptions, such as the electron repulsion [3] or the dispersion force [5]. Moreover, Eq. 35 indicates that $L_e$ decreases with increasing temperature. This is consistent with the expectation that faster particles have a shorter accessible distance, which is vital to the chemical reaction.

**4) Isothermal volumetric density force in the liquid or solid**

$$F_{exo} = kT\frac{dlnV_L}{dr}\vec{r} = \frac{3kT}{r}\vec{r} \tag{36}$$

It is a rotational force similar to the gas phase. Although it may explain some properties of a liquid, e.g., a slight increase in the mass density from the surface to the core of a droplet [40], the spontaneous deformation of a liquid droplet into a sphere, etc., it is inadequate to describe the forces in a liquid. The system potential and environmental potential of the fluid molecules are required. Analogous to Eqs. 14 and 15, the exothermic and endothermic equations may be written.

$$Q^L_{exo} = Q^G_{exo} - RTln\frac{V_L}{V_S} - \frac{\pi^\delta mg}{V_L} - \varepsilon \tag{37}$$

$$Q^L_{endo} = Q^G_{endo} - RTln\frac{V_L}{V_S} + \frac{\pi^\delta mg}{V_L} + \varepsilon \tag{38}$$

Since a liquid cannot exist independently of the gas, the exothermic and endothermic equations must be included. $Q^L$ and $Q^G$ refer to the heat of liquid and gas, respectively. $V_S$ is the molar volume of the solid, and $\delta$ is a coefficient. $\varepsilon$ represents other interactions, such as chemical reactions and irradiations. An atom or a molecule is a highly compressed electron-gas system, which will be addressed in the following sections. As shown in Eq. 37, the item in $Q^G{}_{endo}$ may offset $V_L$, and in Eq. 38 as well. This indicates that the liquid phase is an intermediate phase.

Anyway, the thermodynamics of the liquid or solid is another topic not discussed in this paper.

Finally, the heat-transfer velocity can be derived from Eq. 14.

$$\frac{dQ_{exo}}{dt} = \left(1 - 2ln\frac{L}{r_B}\right) \cdot m_A v_1 \vec{a} + \left(\frac{90\pi^{\alpha-1} m_e g}{4N_A L^4} - \frac{3kT}{L}\right) \cdot v_2 + \frac{3kT}{r} \cdot v_r \quad (39)$$

Hence, the power is,

$$\epsilon = \sum F_i \cdot v_i \quad (40)$$

## Extrapolating Thermal Forces to Microscopic Systems

## 1) Electricity and Mass

Electricity is a particular discipline in modern physics in which mutual repulsion between particles is introduced. In contrast, there is only the universal gravitational attractive force in classical mechanics. However, the above analysis has disclosed that mutual repulsion exists in the macroscopic world. Therefore, the particularity of electricity vanishes. It provides a chance to incorporate the electric interactions into the present theory.

Coulomb's law should be correct because it is an observational result. The defect is that charge is not defined. Hence, to incorporate it, we must replace charges with mass. As discussed above, Coulomb's potential is unlikely to come from gravitation, but for now we have no better choice. Substituting the environmental potential in Eq. 14 with $\frac{k_e q^2}{L}$ in vacuum, where $k_e$ is Coulomb's constant and $q$ is the electron charge, we get the following equation in an electron gas system.

$$Q_{exo} = \frac{3}{2}kT_e - RT_e ln\frac{V_G}{V_S} - \frac{k_e q^2}{L} \quad (41)$$

$T_e$ is the electron temperature, and $V_s$ is the molar volume of electrons in a proton. Since $\frac{90\pi^{\alpha-1} m_e g}{4N_A L^4} = \frac{k_e q^2}{L^2}$, where the SI unit of $L$ is $m$, and the electron mass ($m_e$) is $9.11\times10^{-31}$ kg, for two electrons, it is computed.

$$\alpha_{e\leftrightarrow e} = 1.7479 lnL + 42.9355 \quad (42)$$

Similarly, for two protons,

$$\alpha_{p\leftrightarrow p} = 1.7479 lnL + 36.3688 \quad (43)$$

Moreover, the attractive force is $\frac{3kT_e}{L}$. At thermal equilibrium, $\frac{k_e q^2}{L^2} = \frac{3kT_e}{L}$. Hence,

$$T_e = \frac{k_e q^2}{3kL} \quad (44)$$

By Eq. 44, we can compute the equilibrium distance between two electrons or protons.

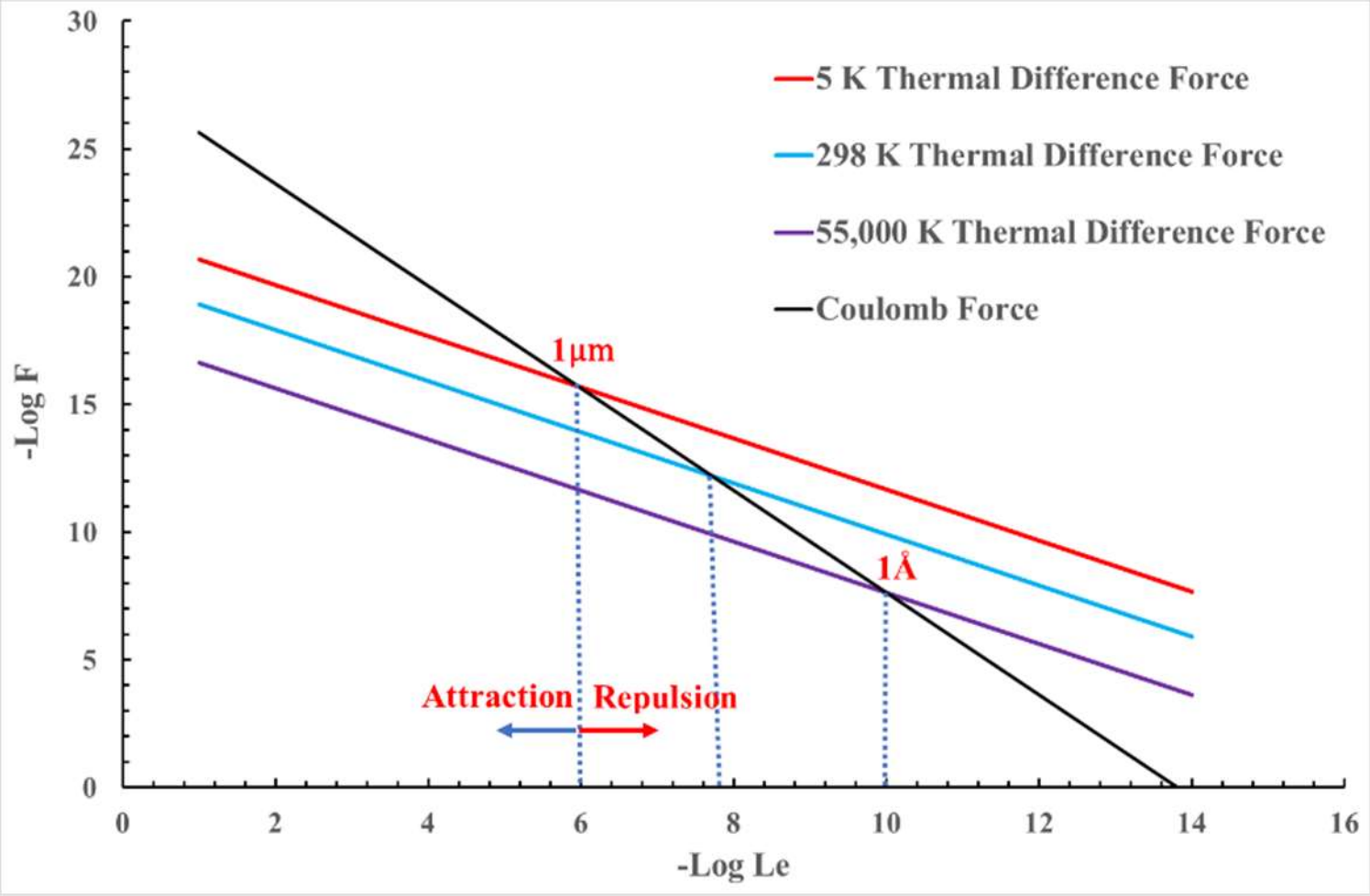

Figure 5, Mutual interactions between two electrons

Fig. 5 exhibits the interactions of two electrons or protons. The cross points indicate the equilibrium distances. As shown in Fig. 5, at 5 K the equilibrium distance is 1 μm; at 298 K, it is about 10 nm; and at $5.5\times10^4$ K, it is 1 Å. This explains why an electron cannot fall into the nucleus or escape the atom. On the other hand, we can now freeze the macroscopic temperature of hydrogen gas with lasers. It can be regarded as the environment shaping the gas molecules by forcing them to condense. As discussed above, this operation simultaneously forces the subsystems, i.e., electrons, to disperse; thus, the electron temperature decreases, and the atom's volume expands. This result can explain many phenomena, such as expanding with heat and contracting with cold, and the fragility of materials at very low temperatures. Superconductivity, superfluidity, and quantum tunneling are also interpretable because, at lower electron temperatures, electrons can leave the nucleus and repel each other at considerable distances. It is well known that He becomes a superfluid near 0 K. The liquid loses viscosity and can creep. This indicates that the attractive force vanishes between two atoms, but increases between helium atoms and the container walls. Certainly, 0 K refers to the temperature of He atoms' thermal motions, rather than the electron temperature.

Some readers may laugh at the mutual attraction of two electrons. We think of point discharge as a typical example where many electrons swarm to the metal tip. We now understand that the electron's temperature on the tip is low because the constriction from surrounding electrons is weak. Moreover, congruent charges coexist on the metal surface rather than being homogeneously distributed throughout the body of materials, e.g., electric capacitors. These examples provide evidence of electron attraction.

Cations and anions are non-equilibrium thermal systems; therefore, they are different from electrons, though they are also "charged." Hence, electricity can be understood in terms of the mass and temperature of electrons and protons.

## Atom Structure

### 1) Structure of a nucleus and magnetism

Coulomb's law states that the force between a neutron and a proton is zero. Hence, to explain the assembling role neutrons play in a nucleus, physicists suggested a model in which a proton induces quadrupole polarization on a neutron surface. It is the unique choice in classical electricity because the repulsion between two protons must be eliminated. However, in the present theory, we have another choice, i.e., the thermal-difference force. Since a proton is equivalent to a neutron in mass, and the nucleus is an adiabatic system, only the kinetic energy and the environmental potential work. Assuming that protons are hot, whereas neutrons are cold, the force that a proton exerts on a neutron is,

$$F_{exo} = -\frac{\sqrt{2}m_p v_0}{4}(1+cos\varphi)^{-\frac{3}{2}}sin\varphi \cdot \frac{d\varphi}{dt} - \frac{k_e q^2}{L_{PN}^2} \tag{45}$$

Using the He nucleus as an example, as shown in Fig. 6, there are two protons exerting force on a neutron. Therefore, the integral force is,

$$F_{exo,int} = -\frac{m_p v_0}{2}(1+cos\varphi)^{-2}sin^2\varphi \cdot \frac{d\varphi}{dt} - \frac{\sqrt{2}k_e q^2}{L_{PN}^2}(1-cos\varphi)^{\frac{1}{2}} \tag{46}$$

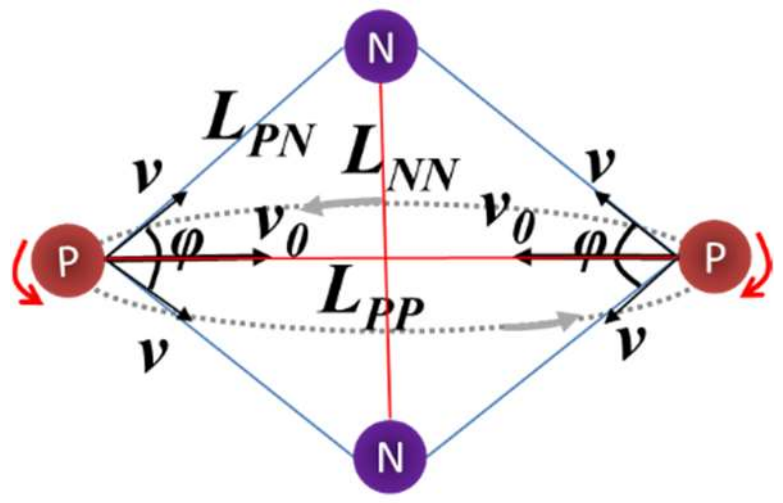

Figure 6, Structure of the nucleus

Since two neutrons are isothermal, and so are two protons, as the forces balance,

$$-\frac{k_e q^2}{L_{NN}^2} = -\frac{m_p v_0}{2}(1+cos\varphi)^{-2}sin^2\varphi \cdot \frac{d\varphi}{dt} - \frac{\sqrt{2}k_e q^2}{L_{PN}^2}\sqrt{1-cos\varphi} \tag{47}$$

and $L_{NN} = 2L_{PN}sin\frac{\varphi}{2}$,

$$\frac{k_e q^2}{L_{NN}^2}\left(1-4\sqrt{2}sin^2\frac{\varphi}{2}\sqrt{1-cos\varphi}\right) = \frac{m_p v_0}{2}(1+cos\varphi)^{-2}sin^2\varphi \cdot \frac{d\varphi}{dt} \tag{48}$$

On the other hand, as two protons attain the forces balance, $\frac{3kT_p}{L_{PP}} = \frac{m_p v_0^2}{L_{PP}} = \frac{k_e q^2}{L_{PP}^2}$, and $L_{NN} = L_{PP}tan\frac{\varphi}{2}$.

Hence,

$$\frac{2\sqrt{k_e}q}{\sqrt{m_p}}\left(1-4\sqrt{2}sin^2\frac{\varphi}{2}\sqrt{1-cos\varphi}\right) = L_{PP}^{\frac{3}{2}}tan^2\frac{\varphi}{2}(1+cos\varphi)^{-2}sin^2\varphi \cdot \frac{d\varphi}{dt} \tag{49}$$

If $L_{PP}$ = 3.356 fm [49] is the diameter of a He nucleus, the angular rate can be depicted over the angle. As shown in Fig. 7, the force balance point is located at $\varphi$ = 1.045 (59.9°). The relative thermal difference between a proton and a neutron is 8.88×10$^8$ K ($T_p$ = 3.32×10$^9$ K).

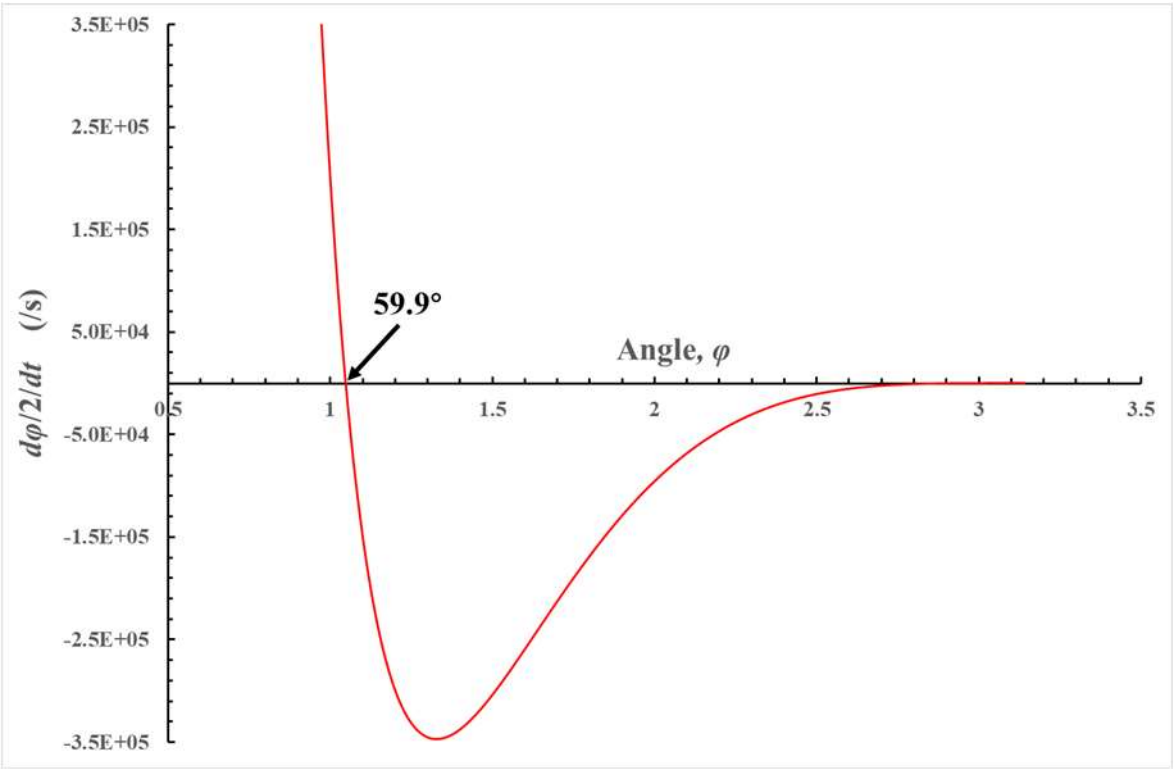

Figure 7, Variation of $\frac{d(\frac{\varphi}{2})}{dt}$ with φ

From the above results, the structure of a He nucleus is $L_{PP}$ = 3.356 fm, and $L_{NN}$ = 1.933 fm in a rhombus with the internal angles of 60° and 120°. A proton carries out a stretching vibration at a balance point with $T_p = 3.32\times10^9$ K, while a neutron conducts a stretching vibration with $T_n = 2.88\times10^9$ K (computed with $L_{NN}$). Hence, the protons spin and rotate around the center of a nucleus, but the neutrons don't spin or rotate due to the balance of two thermal-difference forces.

Given that four ${}^{1}_{1}\mathrm{H}$ atoms fuse to create a ${}^{4}_{2}\mathrm{He}$, the potential energy of four free protons completely converts into kinetic energy. The total kinetic energy is $E_k = 1.55\times10^8$ kJ/mol (1.60 MeV). Isothermal compression of the four H atoms is estimated.

$$Q_{fusion} = -E_k + RT_p ln\frac{V_G}{V_S} + \frac{N_A k_e q^2}{r_H} = 3.9\times10^8 \text{ kJ/mol} \tag{50}$$

$V_G$ is the molar volume of four protons, and $V_S$ is the molar volume of a ${}^{4}_{2}\mathrm{He}$ nucleus. $V_s = \frac{m_p}{4\rho_p} = 1.01\times10^{-18}$L/mol, where $m_p$ is the proton molar mass, 1.01 g/mol; $\rho_p$ is the density of nucleus, $2.5\times10^{17}$ g/L. Provided that the volume of an ${}^{1}_{1}\mathrm{H}$ atom ($r_H$ = 53 pm) is the volume occupied by a proton; the four protons have the same molar volume. $V_G = \frac{4\pi N_A r_H^3}{3} = 3.7535\times10^{-10}$ L/mol. Eq. 45 gives the exothermic heat of $3.9\times10^8$ kJ/mol (~4.05 MeV per ${}^{4}_{2}\mathrm{He}$). It is comparable to the data reported by Wikipedia.

For other nuclei, the geometry may be more complicated. However, the hypothesis that cold neutrons adhere to hot protons via thermal forces is feasible.

Magnetism is a property of the nucleus. The present theory suggests that the magnetism results from the distribution of neutrons and protons on the surface of the nucleus, namely that neutrons and protons dominate half the surface, respectively. An example is the hydrogen element. A hydrogen atom has no magnetism, but a deuterium atom does.

## 2) Ionization Energy

An atom can be regarded as an adiabatic system consisting of the gaseous electrons and a solid nucleus. Therefore, consider only the conversion between kinetic energy and the environmental potential.

$$Q_{exo} = -\frac{3}{2}RT + \frac{N_A k_e q^2}{r} \tag{51}$$

However, once an electron is taken away, an atom becomes a closed system. The system potential must be considered.

Exceptions are the active metal atoms. Since the surface electrons are free in the metal, their atoms cannot be regarded as adiabatic systems. Instead, the metal atoms can be considered as open systems. If taking an electron away is equivalent to condensing an electron, the equation for the isolated system may apply.

Provided that the calculated radii of atoms are reliable [44] and assumed consistent with the isothermal force equilibrium distance between a nucleus and an electron in an atom, the temperature of electrons on the surface can be calculated by using Eq. 44. Here, the nuclear speed is negligible.

Tab. 2 shows the results. The temperature of the shell electrons is $10^4$-$10^5$ K and varies periodically with the number of protons. For example, from Li to Ne, the temperature of the shell electrons increases from $3.33\times10^4$ to $1.463\times10^5$ K, and repeats from Na to Ar. Correspondingly, the velocity of surface electrons is approximately $10^3$ km/s, which is 100 times slower than the speed of light, $3.0 \times 10^5$ km/s.

Table 2, Temperature, Velocity, Kinetic Energy on the Surface of an Atom, and 1st Ionization Energy

| | $r_0$ (pm) | $T$ ×10⁻⁴ (K) | $\bar{v}_e$ ×10⁻³ (km/s) | $\overline{E_k}$ (kJ/mol) | 1st IE (kJ/mol) Ref.* | | | $r_0$ (pm) | $T$ ×10⁻⁴ (K) | $\bar{v}_e$ ×10⁻³ (km/s) | $\overline{E_k}$ (kJ/mol) | 1st IE (kJ/mol) Ref.* | Calc. |
|---|---|---|---|---|---|---|---|---|---|---|---|---|---|
| H | 53 | 10.49 | 2.18 | 1307 | 1312 | | K | 243 | 1.83 | 1.02 | 285 | 419 | 390 |
| He | 31 | 17.93 | 2.85 | 2235 | 2372 | | Ca | 194 | 2.29 | 1.14 | 357 | 590 | 456 |
| Li | 167 | 3.33 | 1.23 | 415 | 520 | | Sc | 184 | 2.87 | 1.17 | 376 | 633 | 456 |
| Be | 112 | 4.96 | 1.50 | 619 | 900 | | Ti | 176 | 3.02 | 1.20 | 394 | 659 | 453 |
| B | 87 | 6.39 | 1.70 | 796 | 801 | | V | 171 | 3.16 | 1.22 | 405 | 651 | 445 |
| C | 67 | 8.30 | 1.94 | 1034 | 1087 | | Cr | 166 | 3.25 | 1.23 | 417 | 653 | 438 |
| N | 56 | 9.93 | 2.12 | 1237 | 1402 | | Mn | 161 | 3.35 | 1.25 | 430 | 717 | 433 |
| O | 48 | 11.58 | 2.29 | 1443 | 1314 | | Fe | 156 | 3.45 | 1.27 | 444 | 763 | 428 |
| F | 42 | 13.24 | 2.45 | 1649 | 1681 | | Co | 152 | 3.56 | 1.29 | 456 | 760 | 423 |
| Ne | 38 | 14.63 | 2.58 | 1823 | 2081 | | Ni | 149 | 3.66 | 1.30 | 465 | 737 | 415 |
| Na | 190 | 2.93 | 1.15 | 365 | 496 | 845 | Cu | 145 | 3.73 | 1.32 | 478 | 746 | 411 |
| Mg | 145 | 3.83 | 1.32 | 478 | 738 | 993 | Zn | 142 | 3.83 | 1.33 | 488 | 906 | 405 |
| Al | 118 | 4.71 | 1.46 | 587 | 578 | 1108 | Ga | 136 | 3.91 | 1.36 | 509 | 579 | 408 |
| Si | 111 | 5.01 | 1.51 | 624 | 787 | 1088 | Ge | 125 | 4.09 | 1.42 | 554 | 762 | 427 |
| P | 98 | 5.67 | 1.61 | 707 | 1012 | 1138 | As | 114 | 4.45 | 1.49 | 608 | 947 | 450 |
| S | 88 | 6.32 | 1.69 | 787 | 1000 | 1178 | Se | 103 | 4.88 | 1.57 | 673 | 941 | 480 |
| Cl | 79 | 7.04 | 1.79 | 877 | 1251 | 1224 | Br | 94 | 5.40 | 1.64 | 737 | 1140 | 507 |
| Ar | 71 | 10.49 | 1.89 | 976 | 1521 | 1274 | Kr | 88 | 5.91 | 1.69 | 787 | 1351 | 523 |

* The first ionization energy of elements. https://pubchem.ncbi.nlm.nih.gov/ptable/ionization-energy/.

As shown in Tab. 2, the 1st ionization energy (IE) of atoms from H to Ne fits quite well. To account for the IE of elemental atoms, a specific property must be assigned to the environmental potential energy: the force a nucleus exerts on electrons is shared equally among all electrons. For example, there are $i$ electrons ($i = 0, 1, 2 \ldots n\text{-}1, n$) in an atom. As the number of electrons decreases, the potential energy shared by each electron increases.

$$\frac{3kT_i}{r_i} = \frac{8.99\times10^9 nQ^2}{(n-i)r_i^2} \tag{52}$$

$$r_i = \sqrt{\frac{n-i}{n}}r_0 \tag{53}$$

$$T_i = 5.559 \times 10^{-6} \cdot \frac{1}{r_0}\sqrt{\frac{n}{n-i}} = \sqrt{\frac{n}{n-i}}T_0 \tag{54}$$

$$E_{k,i} = \frac{3N_A kT_i}{2} = \frac{3R}{2} \cdot \sqrt{\frac{n}{n-i}}T_0 \tag{55}$$

$T_i$ is the temperature of the remaining $n$-$i$ electrons when $i$ electrons have been removed from $n$ electrons. $n$ is the total number of protons in the nucleus, and also the total number of electrons in an atom. $E_{k,i}$ is correspondingly the molar kinetic energy of an electron. The effect of interactions between two electrons on the surface may be negligible because the mass of a nucleus is much larger than that of an electron.

Accordingly, the $i^{th}$ IE should be, considering that taking away an electron is equivalent to condensing an electron,

$$Q_{exo,i} = \frac{1}{n-i}\left(-\frac{3}{2}RT_i + RT_i ln\frac{V_{G,i}}{V_S} + \frac{N_A k_e Q^2}{r_i}\right) \quad (56)$$

Accordingly, for an H-1 atom ($r_0$ = 53 pm), $T_0$ = 1.05×10$^5$ K, and then $E_{k,0}$ = 1.307×10$^3$ kJ/mol or 13.543 eV. As shown in Tab. 2, the IE recommended by NIST is 1.312×10$^3$ kJ/mol or 13.599 eV. It indicates that the $1^{st}$ IE is the kinetic energy of an electron. Moreover, for a He-4 atom ($r_0$ = 31 pm, $n$ = 2), $T_0$=1.793×10$^5$ K. Therefore, the first electron gives $E_{k,0}$ = 2.238×10$^3$ kJ/mol or 23.164 eV, and the second electron exhibits $E_{k,1}$=4.476×10$^3$ kJ/mol or 46.328 eV. The NIST IEs are 2.37×10$^3$ kJ/mol for the 1$^{st}$ electron and 5.24×10$^3$ kJ/mol for the 2$^{nd}$ electron, respectively. They are also generally consistent. These results indicate that Eq. 55 is invalid for the IE calculations of H and He atoms.

However, for a Li-3 atom ($r_0$ =167 pm, $n$=3), $T_0$=1.00×10$^5$ K. $E_{k,0}$ = 414.667 kJ/mol for the first electron, whereas for the 2$^{nd}$ and 3$^{rd}$ electrons, they are 829.333 and 1244 kJ/mol, respectively. Compared with the NIST IEs for the three electrons (520, 7.298×10$^3$, and 1.182×10$^4$ kJ/mol sequentially), the 1$^{st}$ IE is comparable, but the 2$^{nd}$ and 3$^{rd}$ considerably deviate from their kinetic energies. Applying Eq. 56, the IEs are sequentially 3.21×10$^3$, 1.065×10$^4$, and 1.786×10$^4$ kJ/mol. The 2$^{nd}$ and 3$^{rd}$ IEs fit the reference data well. However, the 2$^{nd}$ IEs of atoms from Be to Ne once again fit the $E_k$. For example, the $E_k$s of the 2$^{nd}$ electron sequentially are 1.429, 1.992, 2.776, 3.538, 4.367, 5.252, and 6.081×10$^3$ kJ/mol, corresponding to the 2$^{nd}$ IEs of reference: 1.757, 2.427, 2.353, 2.856, 3.388, 3.374, and 3.952×10$^3$ kJ/mol. Moreover, the last two IEs are special. The IEs computed by omitting the $\frac{1}{n-i}$ in Eq. 56 fit the reference data. Therefore, Eq. 56 is applicable to the mid-IEs of elements from Be to Ne. Fig. 8 shows the computed results.

As shown in Fig. 8A, the last two IEs are particularly high compared with the others. Remembering that in Eq. 49, the electrons are not distinguishable, this result indicates that the energy of the last two electrons (1$s$ orbit) is special relative to the other electrons (2$s$ and 2$p$ orbits). In contrast, the difference in energy of electrons in the 2$s$ and 2$p$ orbits is insignificant. The calculated IEs are magnified and shown in Fig. 10B. It is subtle that IEs of Ne 2$s$ exhibit, to some extent, a difference from IEs of Ne 2$p$ in the energy level and rate. Ne 2$p$ IEs are smaller and increase slowly as the sequential number increases. In addition, the reference data strongly suggest that electrons with the same sequential number have similar IEs, regardless of the atom type. For example, the 9$^{th}$ IE of F and the 9$^{th}$ IE of Ne, as well as the 8$^{th}$ IE of O and the 8$^{th}$ IE of F, are approximately equal.

IEs of elements from Na to Ar are also calculated and shown in Fig. 9. It is observed that, compared with the reference data, the computed IEs are considerably smaller.

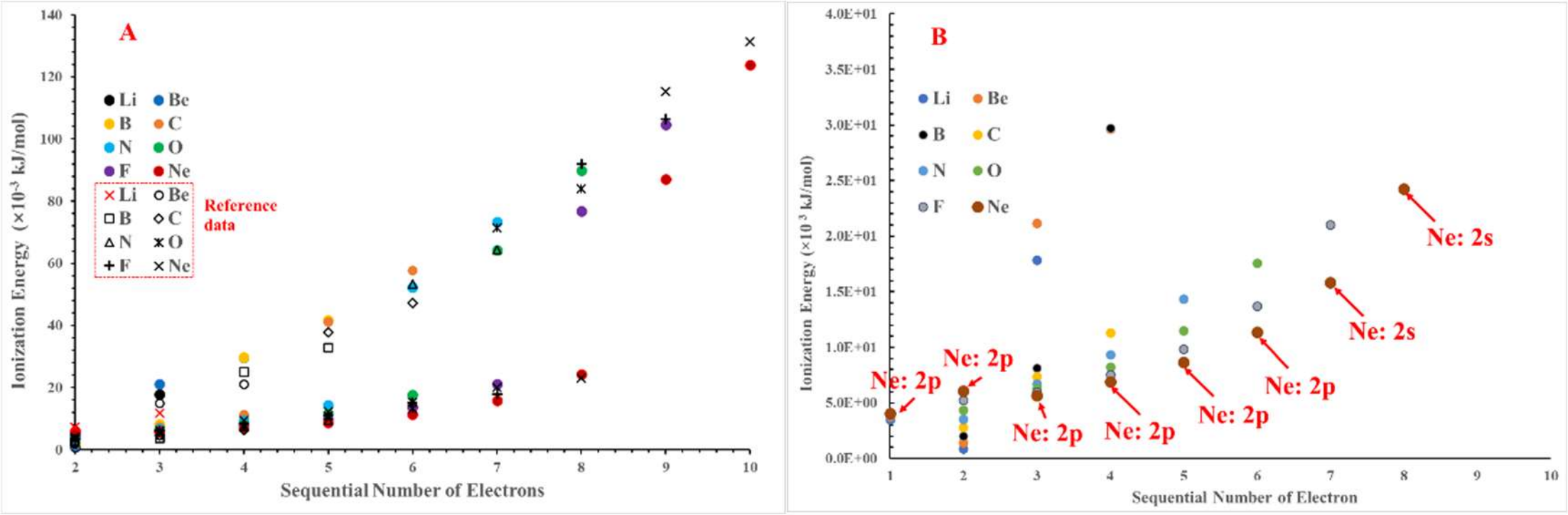

Figure 8, Calculated ionization energies of elements from Li to Ne

(Reference data cited from https://en.wikipedia.org/wiki/Ionization_energies_of_the_elements_(data_page))

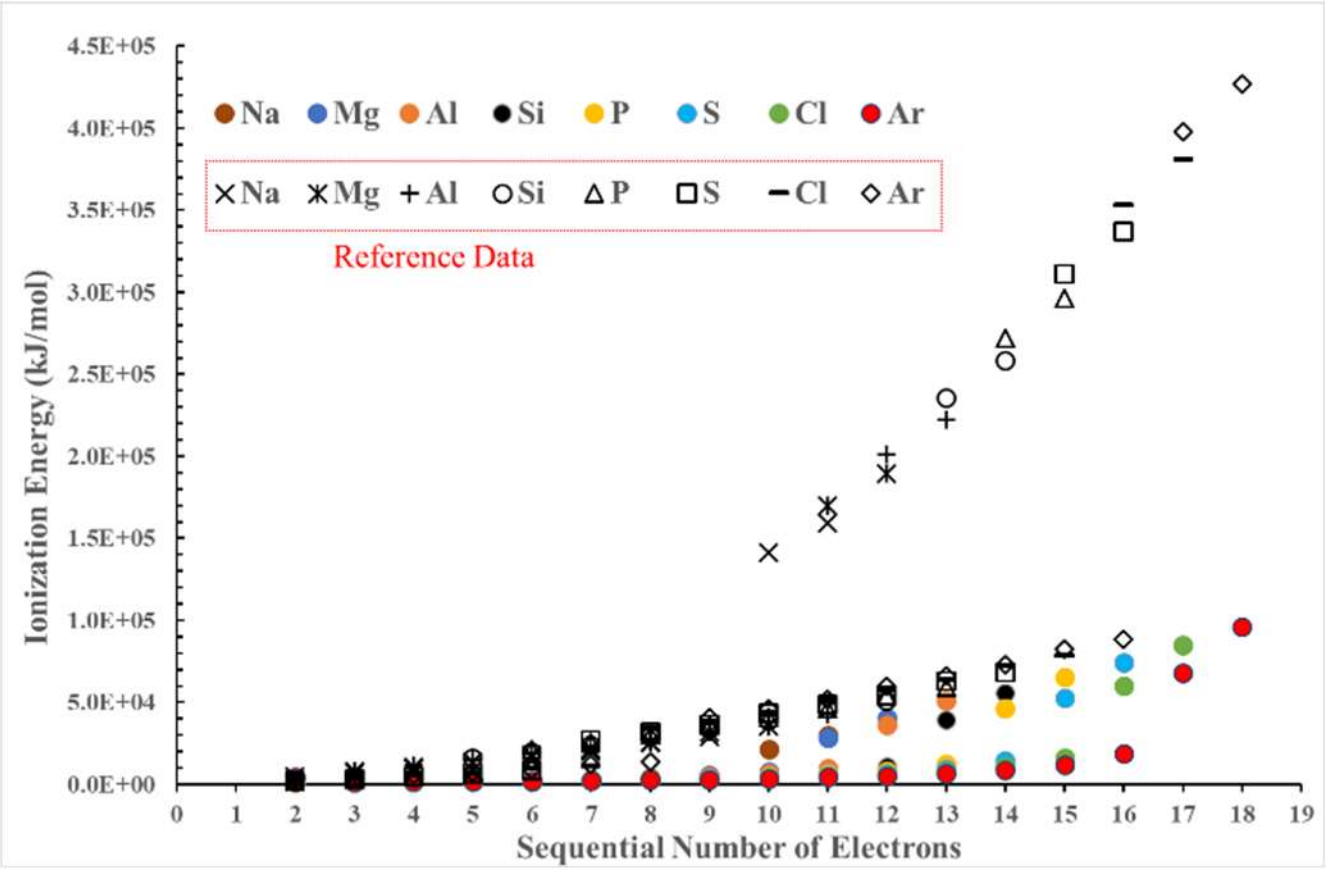

Figure 9, Comparison of the calculated and reference IEs of elements from Na to Ar

The gap between the adjacent IEs can be calculated as,

$$\frac{Q_{exo,i+1}-Q_{exo,i}}{N_A\sqrt{n}} = \left(-\frac{3kT_0}{2}+\frac{k_eQ^2}{r_0}\right)\left[(n-i-1)^{-\frac{3}{2}}-(n-i)^{-\frac{3}{2}}\right]+kT_0\left[(n-i-1)^{-\frac{3}{2}}ln\frac{(n-i-1)^2V_{G,0}}{n^3V_S}-(n-i)^{-\frac{3}{2}}ln\frac{(n-i)^2V_{G,0}}{n^3V_S}\right] \quad (57)$$

The energy gaps of Ar are shown in Fig. 10. The reference data exhibit the most significant gap of 1*s*-2*s*, indicating that the energy level of the 1*s* orbital is much higher than that of the 2*s*. It is consistent with the computed result. As shown in Fig. 10B, the reference shows a second large gap at 2p-3s. In contrast, the gaps between 2s and 2p, and between 3s and 3p, are insignificant. An apparent stepwise energy gap was also observed in the reference data. For example, the gap of 1*s*-1*s* is higher than that of 2*s*-2*s* and 2*p*-2*p*, while 2*p*-2*p* is higher than 3*p*-3*p* and 3*s*-3*s*. This stepwise gap is unclear in the computed data. The reason is that, in the above calculations, an atom is treated as a random system of electrons, like a gas. Therefore, the gap distance between two adjacent layers (e.g., 1*s*-2*s*, 2*s*-2*p*, 2*p*-3*s*, etc.) should be inserted into Eq. 57 to enlarge the temperature differences between layers. However, the gap diagram indicates that the energy difference between 2s-2p and 3s-3p is insignificant.

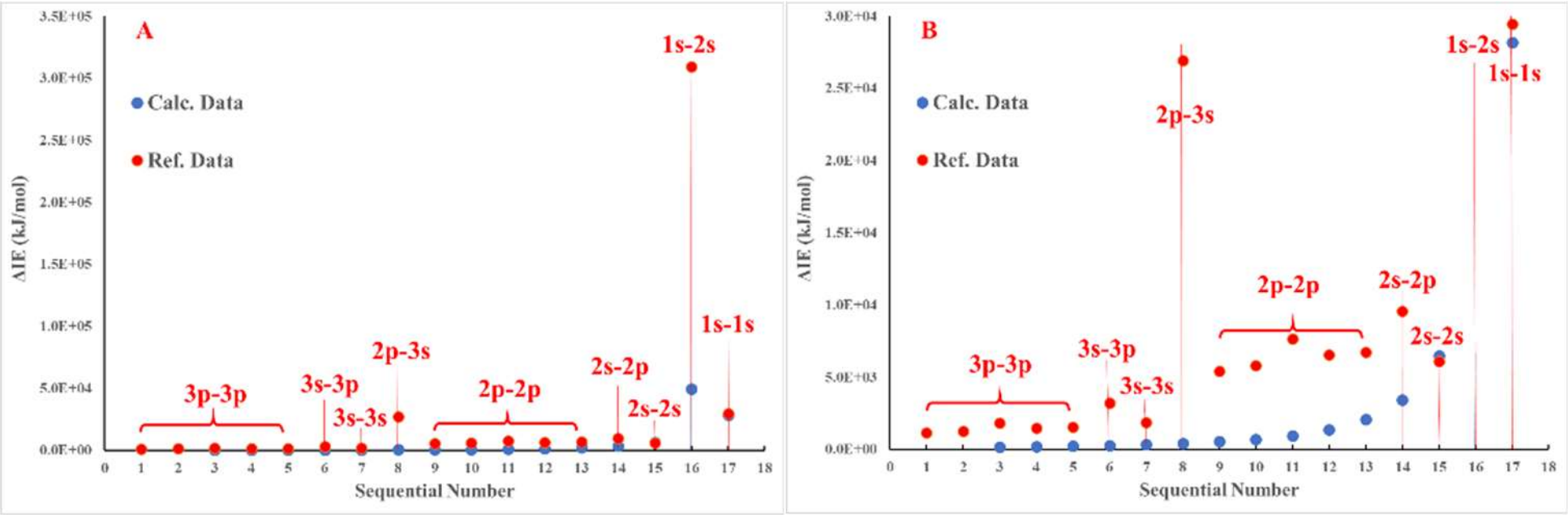

Figure 10, Energy gaps between two adjacent electrons of an argon atom

In summary, the thermodynamic model of an atom appears feasible for describing electron energy levels, as it has produced some positive results. However, the calculations depend heavily on the atomic radii. Many kinds of radii are estimated using different methods, e.g., van der Waals, empirical, and single covalent radii. They are significantly different. For example, the empirical radius of a hydrogen atom is 25 pm, but the van der Waals's radius is 120 pm. This paper employs radii calculated using self-consistent-field (SCF) methods [44].

### 3) Electron orbitals

Based on the above analysis, we can depict an atom's structure. As discussed in the section on Analysis of thermal forces, 3) Volumetric density force: when two isothermal electrons access each other, the volumetric density force makes one electron rotate around the other one while the other one spins. As shown in Fig. 11, since two electrons are equivalent, once force balance is achieved, they orbit each other in a cycle while spinning in opposite directions, i.e., paired. A third electron cannot group with the two electrons because the force balance cannot be attained among three electrons. The diameter of the cycle is the equilibrium distance $L_e$. In an atom, since the force a nucleus exerts on each electron is equal, the paired electrons rotate around the nucleus on an isothermal spherical surface. It resembles a spinning ring, floating in the nucleus. Here, it should be stressed that the temperature of the nucleus, i.e., the kinetic energy, is equal to that of the two electrons. That is to say, the nucleus exerts the volumetric density force on the two electrons. However, since the nuclear mass is much larger than that of an electron, its velocity is negligible. In addition, since the number of protons equals the number of electrons in an atom, the electron-proton interaction is treated as one-to-one and equivalent to the electron-electron interaction, according to Coulomb's law, $L_e = r$. It is a consequence of the isotherm assumption because the radius $r$ determines a flying spin-ring's temperature or velocity ($v_e$). Therefore, the structure of hydrogen, helium, and lithium atoms is determined. As shown in Fig. 11, a He atom resembles a ball (nucleus) adorned with a collar-like ribbon (the 1$s$ orbital of two electrons). It is consistent with the analysis of molecular symmetry [32]. Moreover, since $L_e$=$r$, another parallel collar ribbon is inhibited, let alone the vertical collar ribbon. Furthermore, the second, separated, and parallel collar ribbon (2s) is theoretically admitted. However, the 2$s$ electrons will interact with the 1$s$ electrons to satisfy the requirement of the isotherm. At the same time, 2$s$ electrons must interact with the nucleus along the isotherm. It is impossible because the radius determines the temperature. Therefore, utilizing the dependence of repulsion-attraction on the distance, a non-isothermal orbit should be the best choice for the 2$s$ electrons, namely that the two paired electrons are isothermal but not the nucleus. It can be applied to the remaining electrons.

If four hydrogen atoms combine to form a helium atom, two electrons are released to decrease the temperature of the two protons. The remaining two electrons are paired and attain the isothermal condition. The paired electrons simultaneously experience a force from the nucleus, and then acquire a rotational velocity equivalent in magnitude to the velocity of the paired electrons. Since an electron accepts three forces, i.e., 2 protons and 1 electron, $T = \sqrt{3}T_0$, further, $r = \frac{\sqrt{3}}{3}r_0$, where $T_0$ and $r_0$ are the electron temperature and radius of a hydrogen atom, respectively. Meanwhile, the circular motion becomes oscillatory because two electrons must move forward at the same velocity. If $r_0$ = 53 pm for a hydrogen atom, $r$ = 30.6 pm for a helium atom (Ref. $r$ = 31 pm). It predicts the helium atom radius very accurately.

When further adding an electron, i.e., Li atom, each 1$s$ electron receives 5 forces (2 electrons and 3 protons), $T_{1s} = \sqrt{5}T_0$, and $r_{1s} = \frac{\sqrt{5}}{5}r_0 = 23.7$ pm, while one 2$s$ electron receives 3 forces (two 1$s$ electrons and a proton, the other two protons are shielded); therefore, $T_{2s} = \sqrt{3}T_0$, and $r_{2s} = \frac{\sqrt{3}}{3} =$ 30.6 pm. It means that the 2$s$ electron experiences the strongest forces as it moves along its orbit, parallel to the 1$s$ orbit. At the top or bottom of the exposed sphere, the forces decay due to the increase in distance from the 1$s$ electrons. In quantum theory, all $s$ orbitals are considered as a sphere because of the uncertainty of electron position. However, in this paper, the certainty of electrons' motion is definite, although measuring their position may be challenging. Hence, using the Li atom radius of 167 pm, the 2s electron should adopt a non-spherical orbit perpendicular to the 1$s$ orbit. From a symmetry perspective, the dumbbell-like orbit is recommended. However, since the electron's velocity is ~$10^4$ km/s and the 2p electrons can accommodate it, an asymmetric orbit may be realistic. Unfortunately, because angular momentum conservation does not apply, we cannot estimate the 167 pm radius.

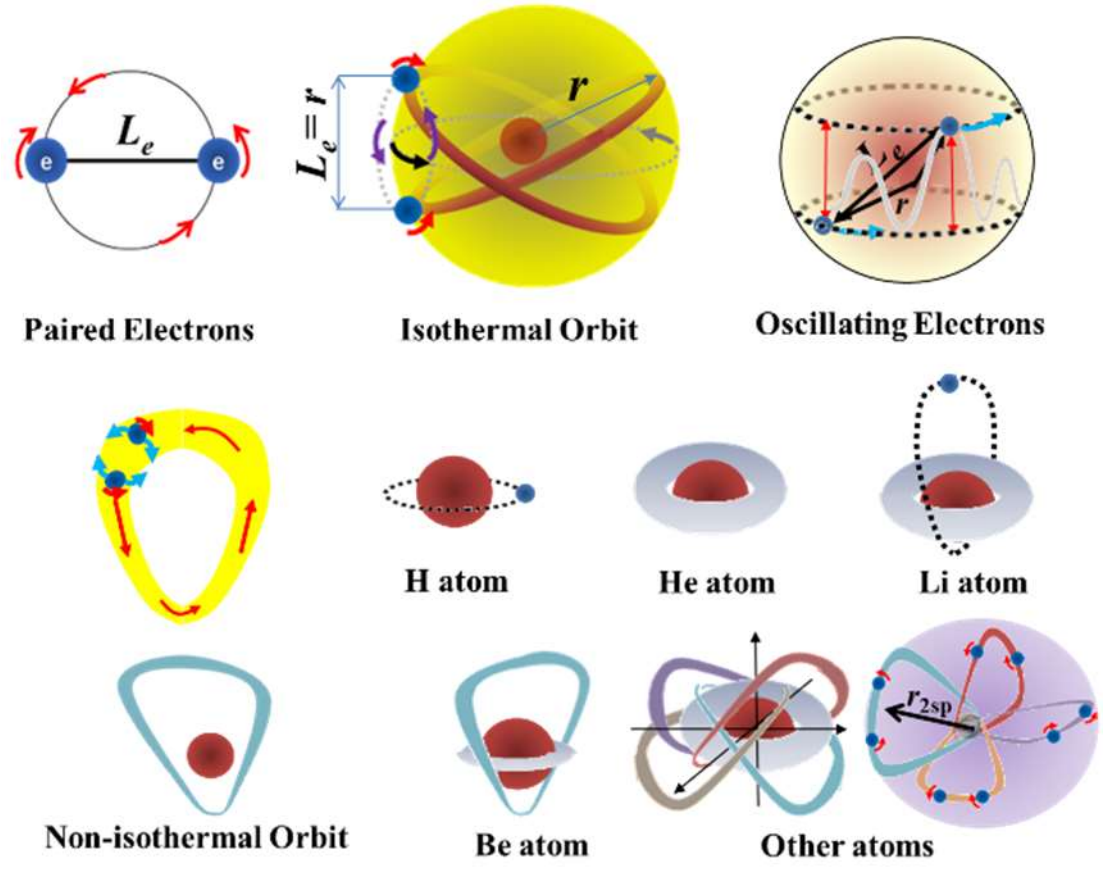

Figure 11, Schematic structure of an atom

For a Be atom, the 1$s$ orbit is further compressed by seven forces; $T_{1s} = \sqrt{7}T_0$, and $r_{1s} = 20.0$ pm, while each 2$s$ electron receives five forces (two protons are shielded); therefore, $T_{2s} = \sqrt{5}T_0$, and $r_{2s} = 23.7$ pm.

Afterward, if we consider 2$s$ as an independent orbital, adding an electron and a proton, each 1s electron is affected by only one force from a proton. $T_{1s} = \sqrt{7+i}T_0$, where $i$ is the number of added protons. The 2$s$ orbit will shield the added electrons. However, according to the reference IE in Fig. 13, the most significant gaps are 1$s$-2$s$ and 2$p$-3$s$, in contrast, 2$s$-2$p$ and 3$s$-3$p$ are insignificant relative to 2$p$-2$p$, 2$s$-2$s$, 3$p$-3$p$, and 3$s$-3$s$. This implies that the 2$s$ and 2$p$ orbitals are distinguishable in IE. They are all non-isothermal orbitals that form a tetrahedron.

Therefore, adding each electron to the 2$s$ orbital, $T_{1s} = \sqrt{7+2i}T_0$, greatly enlarges the energy gap between the 1$s$ and 2$s$ orbitals. On the 2$s$ and 2$p$ orbits, $T_{2s} = \sqrt{9+i}T_0$. Similarly, in the 3$sp$ orbits, the 2$sp$ electrons shield the 1$s$ orbit and all protons. $T_{1s} = \sqrt{19+i}T_0$, $T_{2sp} = \sqrt{19+2i}T_0$, and $T_{3sp} = \sqrt{8+2i-1}T_0$.

As a summary, for H to Ne, $T_{1s} = \sqrt{2n-1}T_0$, $T_{2sp} = \sqrt{2n-3}T_0$, whereas for Na to Ar, $T_{1s} = \sqrt{2n}T_0$, $T_{2sp} = \sqrt{2n-1}T_0$, and $T_{3sp} = \sqrt{2n-13}T_0$, where $n$ is the number of protons.

Therefore, ionization energy may be revised inversely. Taking away an electron doesn't affect the number of protons. However, no tool is currently available to compute the larger radius.

## Chemical bonds

As discussed above, paired electrons are superior to a single electron in terms of energy. Paired electrons give a shorter radius and higher kinetic energy. Although it coincides with the Pauli exclusion principle, here is the consequence of force analysis rather than an artificial assumption.

However, as two H atoms approach each other, the equilibrium distance is determined by the two nuclei because of their vast masses relative to electrons. Since the Coulomb force of proton-proton equals that of proton-electron, the equilibrium distance equals the radius of a H atom. Moreover, the velocities of the two electrons are the same. At the isothermal condition, as shown in Fig. 12, the distance between two nuclei must equal that between the paired electrons and nuclei. Therefore, with two nuclei rotating, a sphere forms with the radius $\frac{\sqrt{3}}{2}r_0$. $\sqrt{3}r_0$ is defined as the bond length.

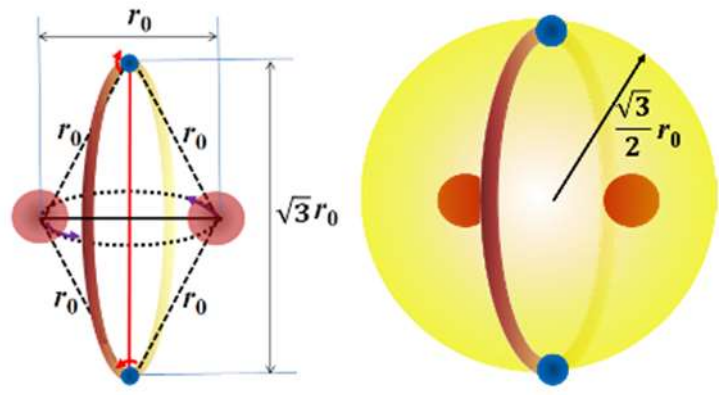

Figure 12, Schematic orbits of an $H_2$ molecule

Hence, the temperature rises to $\frac{2}{\sqrt{3}}T_0$ where $T_0$ is the initial temperature of an electron in a hydrogen atom. The bond energy is defined as the difference in the atoms' kinetic energies.

$$\Delta E = 2 \cdot \frac{3k}{2}\left(\frac{2}{\sqrt{3}} - 1\right)T_0 \tag{58}$$

As shown in Tab. 2, $T_0$=1.049×10$^5$ K, $\Delta E$=403 kJ/mol. This is consistent with the reference value of 432 kJ/mol. [45] The reason is that, as shown in Fig. 12, the $V_G$ in each atom remains when two atoms merge into one molecule. Hence, in Eq. 44, only the kinetic energy and environmental potential energy should be considered. As discussed in the new thermodynamics [29], the environmental potential energy is negligible until the system is subjected to vigorous compression. For an atom, ionization is a kind of compression. As electrons are removed, their temperature increases. Therefore, the environmental potential energy gradually increases. At high ionization, it dominates the exotherm.

According to Eq. 52, the 1*s* orbital electron temperature $T_{1s}$ can be calculated.

$$T_{1s} = \frac{8.99\times10^9 Q^2}{3k} \cdot \frac{n}{2r_{1s}} \tag{59}$$

$r_{1s}$ is the radius of 1s orbit. Using the O atom as an example, since the 6 2*sp* electrons are non-isothermal, $r_{1s}$ may be estimated by Eq. 59.

$$r_{1s} = \sqrt{\frac{2}{n}} \cdot r_0 \tag{60}$$

Therefore, $r_{1s}$ is approximately 24 pm, and T1s is approximately 2.3 × 10$^5$ K.

Apart from the two electrons paired in the 1s orbit, each oxygen atom has four electrons distributed in four non-isothermal orbits, two of which are unpaired. The molecular orbitals are shown in Fig. 16. Two single electrons from two O atoms or one O atom may be paired because the energies of the orbitals are equal. Therefore, $O_3$ molecule can form, and theoretically, so can carbon atoms, forming an oxygen chain. Similarly, for $N_2$, there are three unpaired electrons. As shown in Fig. 13, the bond status may vary between three non-isothermal orbits and one isothermal orbit because all the initial temperatures of the electrons are equal. The homo-pair and hetero-pair of electrons are indistinguishable. Therefore, an oscillating structure is a rational proposal. This means that the bond energy should be close to the single-bond energy rather than the total energy of three single bonds.

Two isothermal electrons constitute the bonds as mentioned above. For two non-isothermal electrons, using $H_2O$ as an example, the radius of an H atom is 53 pm, longer than that of oxygen, 48 pm. Meanwhile, the temperature of an H electron is $T_0$=1.049×10$^5$ K, whereas, for an O atom, the temperature of surface electrons is $T_0$=1.158×10$^5$ K (Tab. 2). When the O-H bond forms, the O electron transfers heat to the H electron by elongating the radius of the new orbit. Since the temperature is inversely proportional to the radius, provided the new radius is averaged,

$$r = \frac{2r_H r_O}{r_H + r_O} = 50.376 \ \text{pm} \tag{61}$$

and then, the bond length is 2r. Meanwhile, Eq. 58 should be revised.

$$\Delta E = \frac{3k}{2}\left(2 \cdot \frac{2}{\sqrt{3}} T - T_{O,0} - T_{H,0}\right) \tag{62}$$

Here, $T_{O,0}$, and $T_{H,0}$ are the initial temperatures of the O and H electrons, respectively.

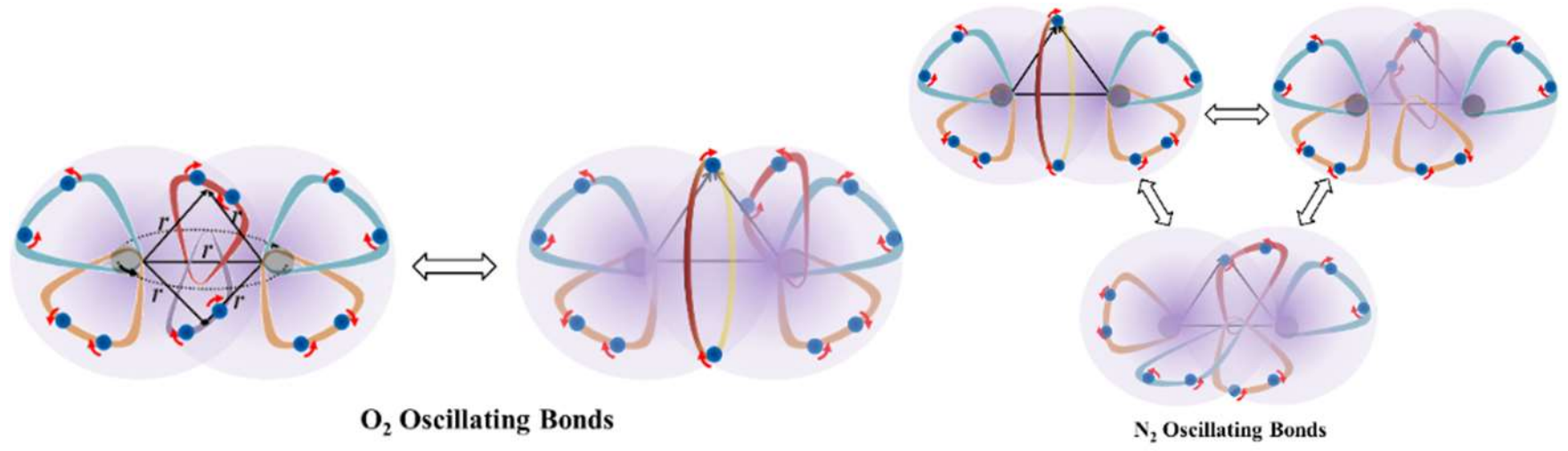

Figure 13, Scheme of bonds in the $O_2$ and $N_2$ molecules

Eq. 62 yields an O-H bond energy of 425.72 kJ/mol, in agreement with the reference data of 423 kJ/mol [45]. Tab. 3 tabulates single-bond lengths and energies for various atomic combinations, along with the reference data from NIST for comparison [44], [45]. However, it must be emphasized that the bond length and energy presented in this paper differ conceptually from those in the reference. The reference [44] defines the bond energy as the enthalpy of bond breaking, which is achieved through various reactions. Different reactions give different values. Hence, the reference data shown in Tab. 3 are arbitrarily selected, unless there is only one, namely that the values are close to those in this paper. Moreover, the bond length [45] is defined as the distance between two nuclei in the reference, whereas in this paper, it refers to the gyration radius of a molecule.

Table 3, Calculated the single bond lengths and energies

| Bond | D[a)] (pm) | Ref. D[b)] (pm) | ΔE$_k$[c)] (kJ/mol) | Ref. ΔH[d)] (kJ/mol) | Bond | D (pm) | Ref. D (pm) | ΔE$_k$ (kJ/mol) | Ref. ΔH (kJ/mol) |
|---|---|---|---|---|---|---|---|---|---|
| H-H | 92 | 74 | 403 | 432 | B-S | 151 | 161 | 245 | 494 |
| B-B | 151 | 159 | 246 | 293 | B-Si | 169 | - | 220 | 285 |
| C-C | 116 | 124 | 320 | 337 | C-N | 106 | 117 | 351 | 305 |
| N-N | 97 | 109 | 382 | 474* | C-O | 97 | 113 | 383 | 358 |
| O-O | 83 | 121 | 448 | 434** | C-F | 89 | 128 | 415 | 439 |
| F-F | 73 | 141 | 510 | 159 | C-Cl | 126 | 165 | 295 | 322 |
| Cl-Cl | 137 | 199 | 271 | 240 | C-Br | 136 | 182 | 274 | 280 |
| Br-Br | 163 | 228 | 228 | 194 | C-I | 147 | - | 253 | 226 |
| I-I | 199 | 266 | 186 | 150 | C-S | 132 | 154 | 282 | 297 |
| S-S | 152 | 189 | 243 | 262 | N-O | 90 | 115 | 414 | 201 |
| P-P | 170 | 189 | 219 | 485 | N-F | 83 | 132 | 446 | 292 |
| Be-H | 125 | 134 | 298 | 222 | N-Cl | 114 | 161 | 327 | 259 |
| C-H | 103 | 112 | 362 | 413 | N-Br | 122 | 177 | 305 | 285 |
| N-H | 94 | 101 | 393 | 389 | O-Li | 129 | 169 | 287 | 331 |
| O-H | 87 | 97 | 425 | 423 | O-F | 77 | 135 | 478 | 155 |
| F-H | 81 | 92 | 457 | 565 | O-Cl | 103 | 160 | 359 | 268 |
| Cl-H | 110 | 127 | 338 | 427 | O-Br | 110 | 172 | 337 | 231 |
| Br-H | 117 | 141 | 316 | 363 | O-I | 117 | 187 | 316 | 180 |
| I-H | 126 | 161 | 295 | 295 | O-Si | 117 | 151 | 320 | 452 |
| Si-H | 124 | 152 | 299 | 310 | O-S | 107 | 148 | 345 | 343 |
| S-H | 115 | 134 | 324 | 340 | O-P | 112 | 148 | 332 | 592 |
| P-H | 119 | 144 | 311 | 322 | O-Na | 133 | 205 | 279 | 322 |
| B-N | 118 | 133 | 314 | 385 | F-Li | 116 | 156 | 319 | 573 |
| B-O | 107 | 121 | 346 | 787 | Cl-Na | 193 | 236 | 192 | 410 |

*Data discrepancy is enormous: 942, 84, 474, 29, and 54 kJ/mol. **Data discrepancy is enormous: 493, 264, 207, and 434 kJ/mol. a) the calculated length of the single bond. b) the bond length in reference [47]; c) the calculated kinetic energy difference; d) the experimental bond energy in reference [46].

As shown in Table 3, a discrepancy is apparent, although most of the data are comparable. For example, the reference length for the $O_2$ molecule is 121 pm [47]. However, in this paper, the radius of an O atom is 48 pm. The bond length of $O_2$ is longer than the two-fold radius of an O atom. It is a widespread phenomenon. Moreover, in this paper, the longer bond corresponds to the lower temperature of paired electrons. For example, the bond lengths of $N_2$ and $O_2$ are 97 pm and 83 pm, corresponding to bond energies of 382 kJ/mol and 448 kJ/mol, respectively. However, the longer bond gives a larger enthalpy in the reference. For example, in reference [46], the bond energies of O-O and F-F are usually averaged to provide 146 and 159 kJ/mol, corresponding to the bond lengths 121 and 141 pm, respectively. This contradicts common sense. Based on the concepts in this paper, these data are not acceptable. Therefore, the closest energies of N-N and O-O are selected for listing in Tab. 3. The F-F value of 159 kJ/mol is also anomalous, but we cannot find a suitable value. Since the radius of an F atom is 42 pm, 141 pm for an F-F bond in the reference is overly exaggerated. In addition, the F-F bond energy is smaller than the Cl-Cl bond energy. Therefore, this data is not recommended as a reference.

On the other hand, the enormous discrepancy in energy appears when multiple bonds possibly exist between two atoms, such as B-O, N-N, O-O, P-P, N-O, O-P, etc., and fluoride compounds, such as F-F, F-H, N-F, etc. When comprehensively comparing the lengths and energies of all reference data, we think these bond energies are not convincing. For example, a B-O length is 121 pm, but the energy is 787 kJ/mol, while the length of O-O is also 121 pm, with a maximum bond energy of 434 kJ/mol. The first question arises from the equal length. It is impossible because the radii of B and O are 87 and 48 pm, respectively. The second question is whether the B-O energy is twice the energy of O-O. We cannot find a rational reason. As for the salts, since the metallic atom has many vacant orbitals, multiple bonds may form. In addition, the electrons in cations are heated, while the electrons in anions are cooled down. Therefore, as in the interaction between a proton and a neutron in a nucleus, the thermal forces are likely dominant. Furthermore, the reference provides data at 0 K to eliminate entropy. However, this paper does not consider the effect of molecular temperature. As discussed in the following section, molecular temperature affects bond energy.

Overall, the computed bond lengths and energies are reasonable compared with the reference data. Therefore, this paper presents a novel, concise method for calculating single-bond strength.

## Hydrogen bonds

Hydrogen bonds (HB) are special bonds formed between two molecules through a hydrogen atom. Fig. 14 exhibits a typical HB forming between two $H_2O$ molecules.

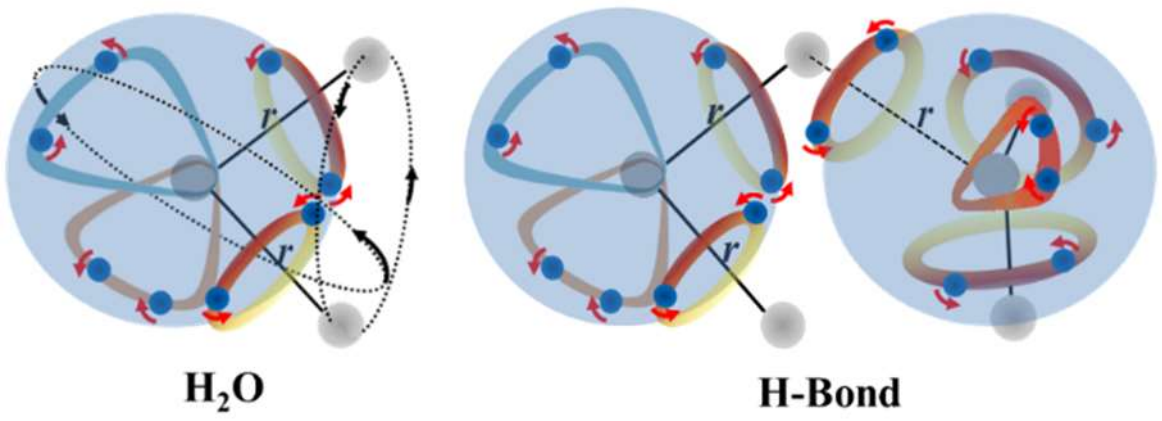

Figure 14, Scheme of water bonds and H-bond between water

As shown in Fig. 14, the HB orbit is isothermal. Since the mass of oxygen is 16 times that of hydrogen, two hydrogen nuclei rotate around the oxygen nucleus while revolving around their common axis. Such rotation of H nuclei exposes the hydrogen nuclei to the outer, or uncovered, region. Hence, HBs can form by accepting two paired electrons from the other O atom. The HB length should equal the O-H single bond length because there is no reason to change it. However, the HB energy differs from the O-H single bond since two electrons are contributed by one orbital of O. Eq. 58 is rewritten as,

$$\Delta E = 2 \cdot \frac{3k}{2}\left(\frac{2}{\sqrt{3}}T - T_{O,0}\right) \quad (62)$$

Therefore, the HB energy of $H_2O$...H-OH is 289 kJ/mol. Similarly, the HB energies of other molecules are computed and presented in Tab. 4. For $NH_3$ and HF, the strengths of HBs are 493 and 114 kJ/mol, respectively. These values are significantly higher than those reported in the references, ranging from 1-2 kJ/mol to 161 kJ/mol (F-H…FH), and varying with temperature. However, this HB bond-strength calculation doesn't account for molecular temperature, which we discuss in the following section. Therefore, it is not proper to compare the reference HB strength with the calculated data. However, the order of calculated HB strengths gives strong support to the following reactions:

$NH_3 + H_2O \rightarrow NH_4 + OH$

$NH_3 + HF \rightarrow NH_4 + F$

$H_2O + HF \rightleftarrows H_3O + F$ (HF is a weak acid)

In contrast, the order of reported HB strengths doesn't support these reactions. For example, the strengths of HO-H…$OH_2$, $H_2N$-H…$NH_3$, and F-H…FH are about 26, 21, and 161 kJ/mol. These data indicate that $H_2F$ should be formed when HF is mixed with $H_2O$ and $NH_3$, while $H_2O + NH_3 \rightleftarrows H_3O + NH_2$.

Table 4, Calculated HB energies

| Type of HB | Single Bond (kJ/mol) | HB Energy (kJ/mol) | Type of HB | Single Bond (kJ/mol) | HB Energy (kJ/mol) |
|---|---|---|---|---|---|
| $H_3N$…H-$NH_2$ | 393 | 493 | $H_3P$…H-$PH_2$ | 311 | 912 |
| $H_2O$…H-OH | 425 | 289 | $H_2S$…H-SH | 324 | 844 |
| HF…H-F | 457 | 114 | HCl…H-Cl | 338 | 768 |

On the other hand, as shown in Tab. 4, the HB energies are considerably larger than their single bond energies, except for the HB bonds of $H_2O$…H-OH and HF…H-F. This means the HB bond forms preferentially over the single bond. Hence, the energy is shared among the four *sp* orbitals. As shown in Fig. 15, the H atoms share their energy with one *sp* orbit of the P atom through interactions. A similar situation occurs in the formation of $H_2O$, $NH_3$, HF, and other molecules through the addition of H atoms. Therefore, the HBs energies should be modified. For example, the temperature of the paired electrons in the *sp* orbitals of $H_2O$ can be estimated with the equations in the section on Atom Structure. It rises from $1.158\times10^5$ K in an O atom to $1.27\times10^5$ K, equivalent to that of O-H. Accordingly, the HB energy decreases from 289 kJ/mol to 0. However, the difference from an O atom is apparent. The two protons are located outside an O atom rather than inside the nucleus, which pulls the two 2*sp* orbitals outward and increases their radii. In addition, the distance between an H proton and two bonded electrons is shorter than that between an H proton and two *sp* electrons. As a result, the energy of HB bonds among $H_2O$ molecules is not zero, but smaller than 289 kJ/mol. Unfortunately, the longest radii of *sp* orbitals cannot be computed at present.

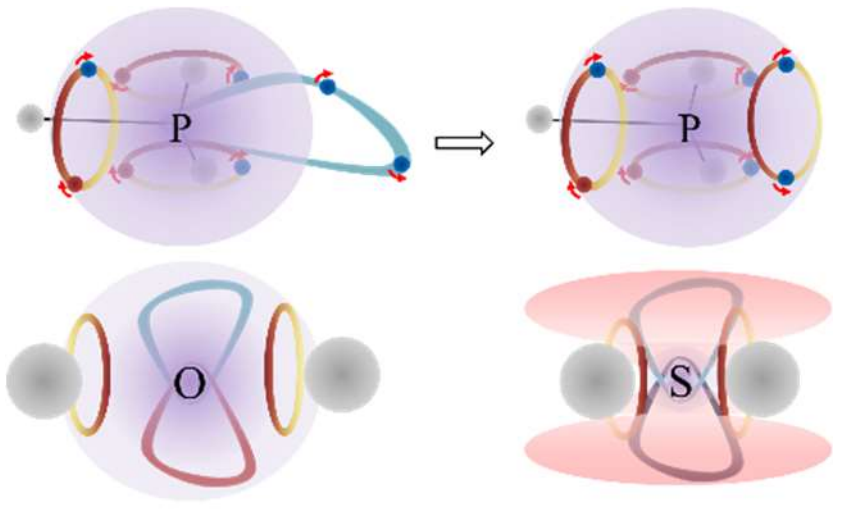

Figure 15, Schematic structure of $PH_3$ molecule and the shape of $H_2O$ and $H_2S$

The energy of the F 2*sp* orbitals is less affected in an HF molecule because there is only one added proton. It is reasonable to infer that the computed HB energy (114 kJ/mol) is consistent with the reference value (161 kJ/mol). In contrast, the 2*sp* orbitals should be severely compressed by the three outside H protons because the three protons are located on one side, opposite to the *sp* orbital. As shown in Fig. 15, the remaining *sp* orbital may convert into an *s* orbital.

However, the most crucial factor is single-bond length. The net lengths of P-H, S-H, and Cl-H are 68, 66, and 63 pm, but the radii of P, S, and Cl 3*sp* orbitals are 98, 88, and 79 pm. The net length of a single bond is shorter than the radius of a 3*sp* orbital. In contrast, the net lengths of N-H, O-H, and F-H are longer than the radii of N, O, and F 2*sp* orbitals. As the molecule rotates, the proton protrudes from the surface of the $NH_3$, $H_2O$, and HF spheres, whereas it concaves inward in the $H_2S$ and HCl spheres. Hence, it is difficult to form HB bonds between $H_2S$ and HCl molecules. However, $PH_3$ does not prefer $PH_4$ in terms of energy, and the same applies to $NH_3$.

## Frequency and Energy

As shown in Fig. 12, both nuclei and electrons are rotating. Hence, the rotating frequency can be calculated. $v = 2\pi r f$ and $\frac{3}{2}kT = \frac{1}{2}mv^2$, $f = \frac{1}{2\pi r}\sqrt{\frac{3kT}{m}}$, where $v$ is the linear velocity of rotation, $m$ is the total mass of a particle, $r$ is the radius, and $f$ is the frequency. Moreover, if the total charge of a particle is $nq$, where $n$ is the number of charges, and force equilibrium is attained, applying Eq. 54,

$$f = \frac{3kT}{5.646\times10^{10}n^2q^2}\sqrt{\frac{3kT}{m}} = \frac{1.843\times10^{-7}}{n^2m^{0.5}} \cdot T^{\frac{3}{2}} \quad (63)$$

Using $H_2$ in the earth as an example, calculate the rotational frequency of two protons with $r = 53$ pm, $n = 1$, and $m_p = 1.6726\times10^{-27}$ kg. $T_p = \frac{8.99\times10^9 Q^2}{3kr} = 1.05 \times 10^5$ K. $f_p = 1.53 \times 10^{14}$ Hz. Similarly, the rotational frequency of two electrons can be calculated with $r = \frac{\sqrt{3}}{2} \times 53$ pm, $n = 1$, and $m_e = 9.11\times10^{-31}$ kg. $T_e = 1.21 \times 10^5$ K. $f_e = 8.32 \times 10^{16}$ Hz.

As we know, there are four characteristic spectral lines in the visible zone emitted from an excited $H_2$, i.e., 656, 486, 434, and 410 nm. Using the speed of light, $c = 3 \times 10^8$ m/s, the corresponding frequencies are 4.57, 6.17, 6.91, and 7.32 × $10^{14}$ Hz. In the ultraviolet zone, the lines are located at 121, 102, and 97 nm, corresponding to frequencies of 2.48, 2.94, and 3.09 × $10^{15}$ Hz. These frequencies are faster than $f_p$, but slower than $f_e$. Moreover, in the infrared red zone, the spectral peak at 1.2-2.5 μm corresponds to 1.2-2.5 × $10^{14}$ Hz, which is close to $f_p$. Considering that the temperature on the Sun is much higher than that on Earth, it implies that those light wavelengths are possibly determined by the rotation of the two nuclei in $H_2$ molecules. It must be pointed out that the wavelength (~0.3 nm) of the electron and proton rotation is several orders of magnitude shorter than the above wavelengths because the light velocity, $c = 3 \times 10^8$ m/s, is about 100 times the electron velocity in $H_2$.

## Velocity of Heat Transfer by Electrons

Fig. 16 shows a hot electron transferring heat to a cold electron located at a distance of $L$.

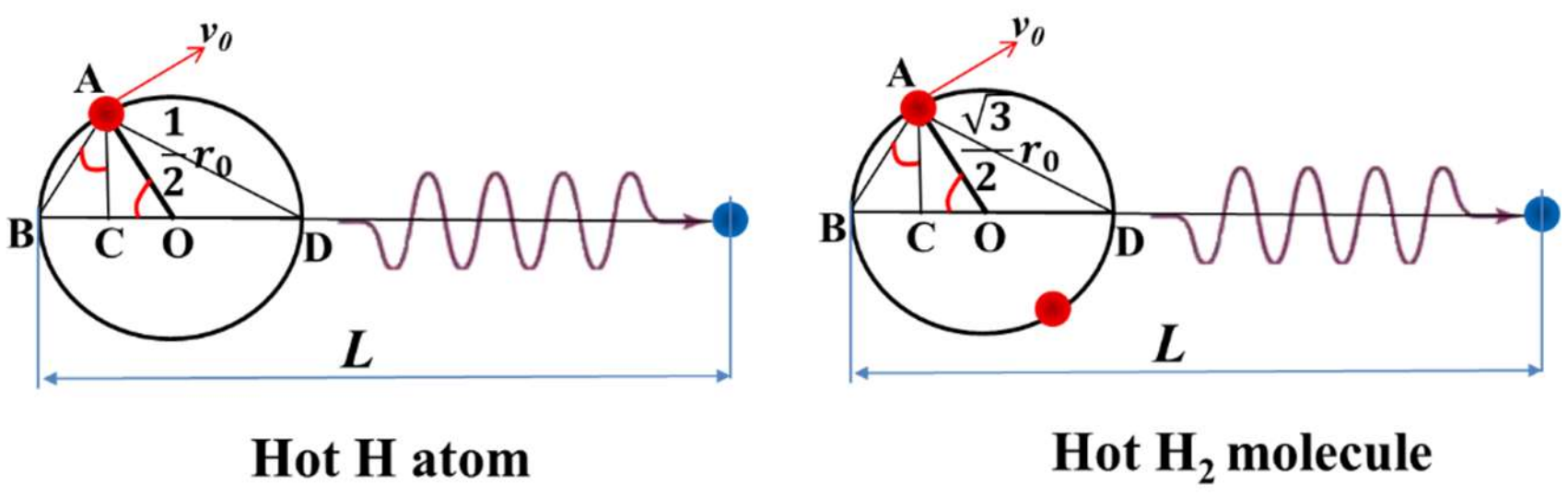

Figure 16, A hot electron transfers heat to a cold electron

In a hydrogen atom, if ∠AOC = $\omega t$, where $\omega$ is the angular rate, and $t$ is time, $BC = r_0 sin^2(\frac{\omega t}{2})$. Since the heat transfer rate is $\frac{dQ_{exo}}{dt} = \frac{dQ_{exo}}{dL} \cdot \frac{dL}{dt} = F_{exo} \cdot \frac{dBC}{dt}$ (Eq. 40), the velocity of heat transfer is

$$v_{exo} = \frac{dBC}{dt} = 4\pi r_0 f_e sin(\pi f_e t)\cos(\pi f_e t) \qquad (64)$$

Strictly speaking, $r_0$ and $f_e$ vary over time during heat transfer because of kinetic energy loss. However, for demonstration purposes, we treat them as constants. Therefore, with $f_e = 1.31\times10^{16}$ Hz,

$$v_{exo} = 8.72 \times 10^6 sin(\pi f_e t)\cos(\pi f_e t) \qquad (65)$$

Moreover, a hot H atom here is actually a normal H atom on Earth. A temperature of close to $1.57\times10^7$ K in the Sun's core and 5800 K on the Sun's surface is reported. Hence, in the Sun, the temperature of electrons must be much higher than that on Earth.

Converting Eq. 65 into a figure, it is shown in Fig. 17. The negative values are omitted because the heat transfer is directed towards the other electrons. The velocity is discontinuous in time. This means that energy is transferred piece by piece, and indeed it is proportional to the wave's amplitude, as a usual wave does. However, as discussed above, the faster the rotation rate of the nuclei, the faster the rate of electrons. Therefore, the power of light is proportional to the frequency of light.

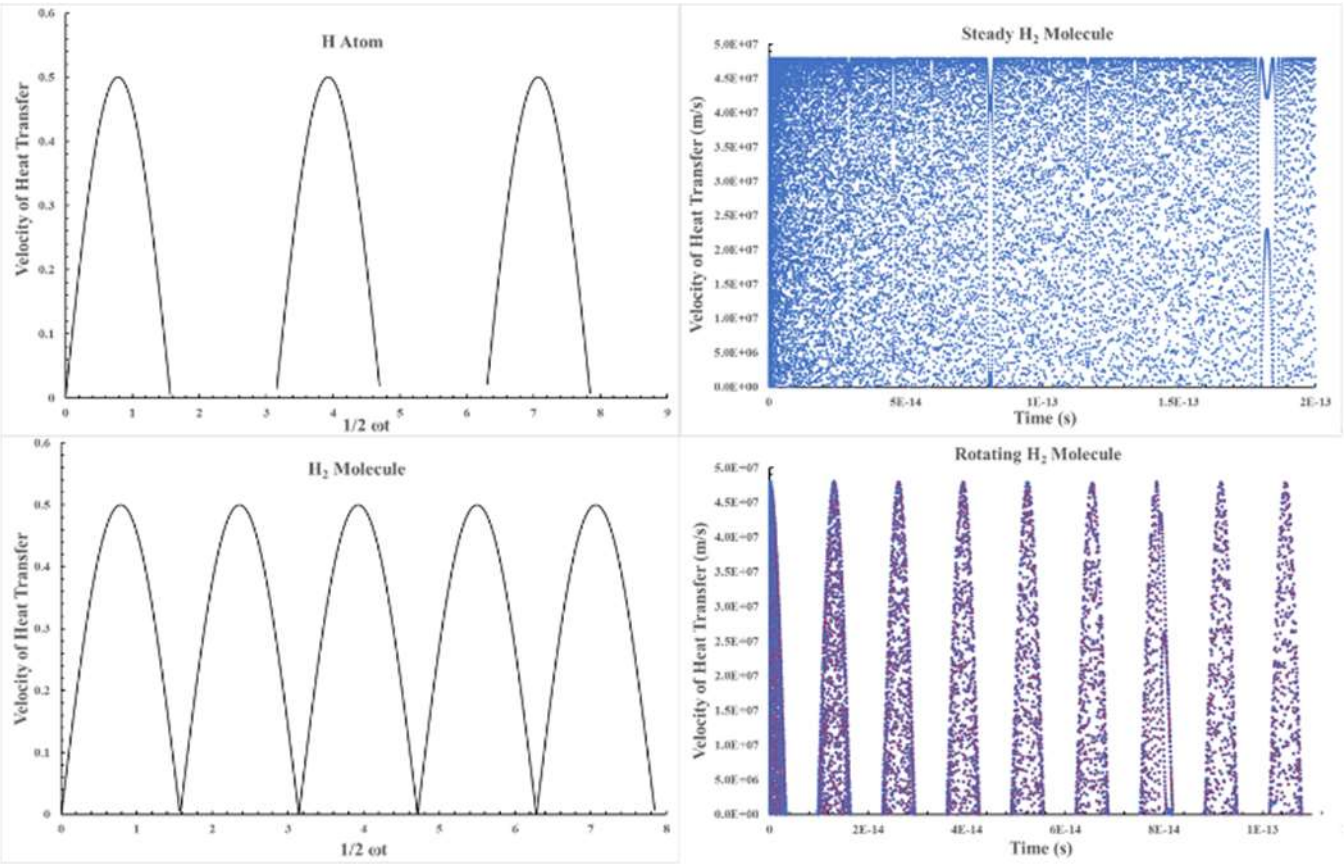

Figure 17, Variation of heat transfer velocity with time

Similarly, we compute the heat-transfer velocity of a normal electron in an $H_2$ molecule. As shown in Fig. 17, the velocity varies periodically and continuously with time. However, when we account for the molecule's rotation (Eq. 66), a discontinuity appears. This means the rotational frequency of an electron determines the heat-transfer velocity, but the rotational frequency of a molecule determines the frequency of the transfer process. It suggests that the speed of light is comparable to that of an electron, while the wavelength of light is similar to the molecule's rotational wavelength. Moreover, the velocity is independent of the distance between two objects. Furthermore, the heat is transferred piece by piece to a cold electron. It is consistent with the quantum theory.

$$v_{exo} = 9.59 \times 10^7 sin(\pi f_e t)\cos(\pi f_e t)\cos(2\pi f_p t) \qquad (66)$$

If a particular factor slows molecular rotation, a red shift is expected. For example, the solid state may slow a molecule's rotation. Compared with the gas state, the solid state may give a longer wavelength.

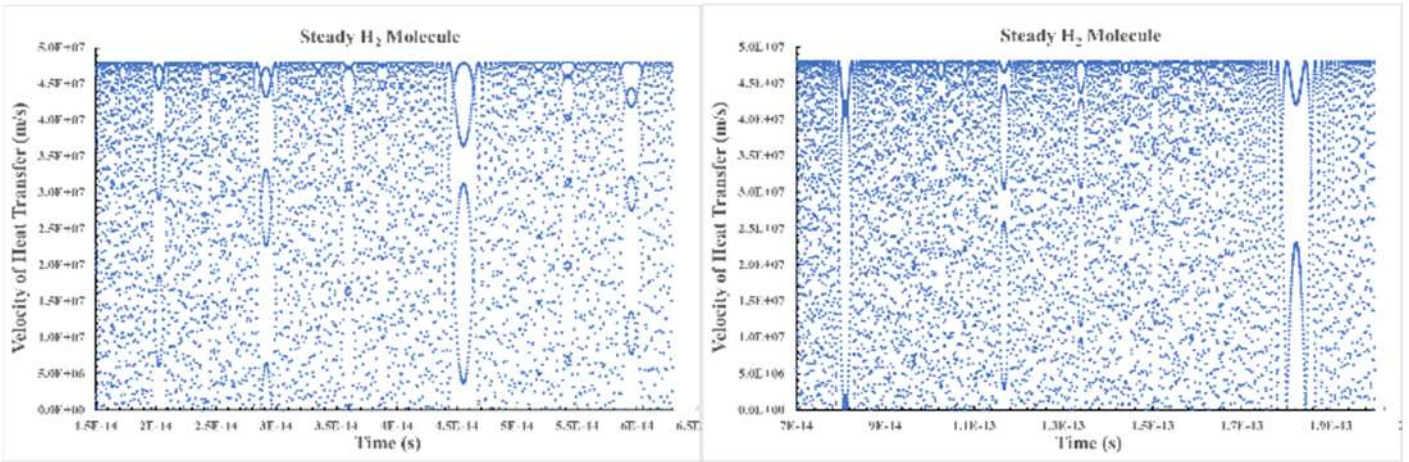

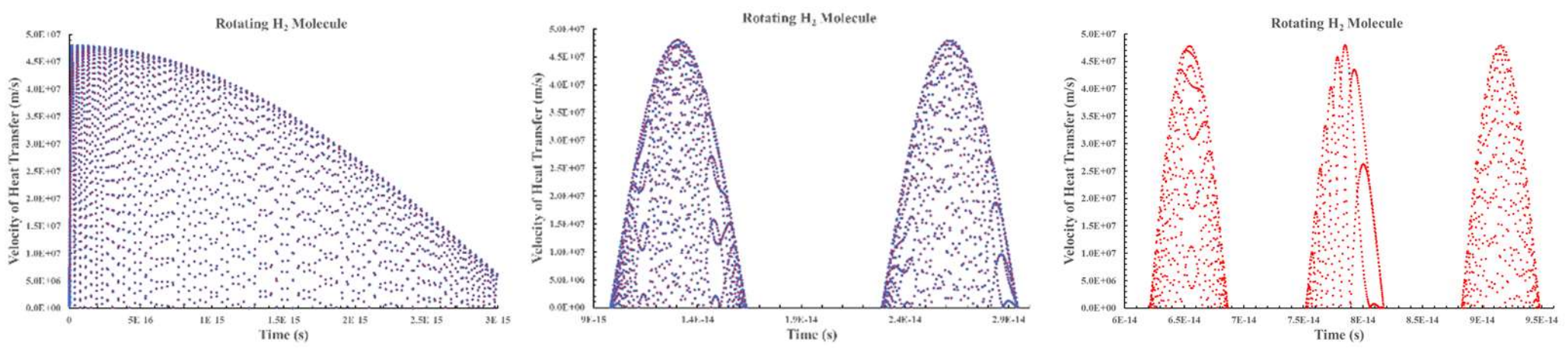

Figure 18, Regular patterns

Anyway, the implication of Fig. 18 is unknown.

The power of heat transfer resulting from the temperature difference is, assuming the temperature of the cold electron is negligible,

$$\epsilon_{exo} = \frac{dQ_{exo}}{dL} v_{exo} = -(m_e \cdot \frac{dv_{exo}}{dt} + \frac{k_e q^2}{L^2}) v_{exo}$$

$$\epsilon_{exo} = -\{2\pi^2 m_e r_0 f_e^2 [1 - 2sin^2(\pi f_e t)] + \frac{k_e q^2}{L^2}\} v_{exo} \quad (67)$$

Taking into account the molecular rotation,

$$\epsilon_{exo} = -\{2\pi^2 m_e r_0 f_e \big[f_e \cos\big(2\pi f_p t\big) - 2 f_e sin^2(\pi f_e t) \cos\big(2\pi f_p t\big) -$$

$$2 f_p \sin(\pi f_e t) \cos(\pi f_e t) \sin\big(2\pi f_p t\big)\big] + \frac{k_e q^2}{L^2}\} v_{exo} \quad (68)$$

Hence, if the distance is adequately long,

$$\epsilon_{exo} = -v_{exo}\{2\pi^2 m_e r_0 f_e \big[f_e \cos\big(2\pi f_p t\big) - 2 f_e sin^2(\pi f_e t) \cos\big(2\pi f_p t\big) -$$

$$2 f_p \sin(\pi f_e t) \cos(\pi f_e t) \sin\big(2\pi f_p t\big)\big]\} \quad (69)$$

It indicates that the distance can be measured based on the light's power. Integrating the power over time yields the distance between the planets, provided that $f_e$ and $f_p$ are given.

However, there are a large number of atoms on the surfaces of both cold and hot objects. As shown in Fig. 19A, a hot molecule can sequentially transfer heat to several cold molecules through rotation. If the hot molecules behind a hot molecule cannot immediately supply the energy, the light inevitably redshifts. Hence, a bundle of light is a mixture of light with various frequencies.

Fig. 19 B shows the mechanism of double-slit interference. As it passes through the slits, the light heats the electrons in the area face-to-face with the slits, and the heated electrons then transfer heat to neighboring electrons. After neighboring electrons transfer heat to the next neighbors, they receive heat again, and the process repeats. Heated neighbor electrons may transfer energy to their neighbors in the opposite direction because heat transfer is random. This produces heat resistance. Hence, heat accumulates in the center band and gradually decays as the distance from the center increases. It resembles the mechanical wave produced by dropping a stone into water. From a probabilistic viewpoint, electrons in some places must remain in an oscillating mode, waiting for the heat balance between the two sides. An observer is always an entity with much colder electrons. It preferentially receives more light energy emitted from the heat-accumulation area. Therefore, the interference pattern is observed. Projecting electrons is similar to light. However, projecting atoms or molecules is another situation. Both atoms and molecules are adiabatic systems. When a hot atom collides with a cold one, electrons and the nucleus of the cold atom are heated. However, due to the large mass of the nucleus,

most energy is possibly emitted in the form of atomic vibration. The energy is low. Hence, an observer with very cold electrons is required to observe the interference.

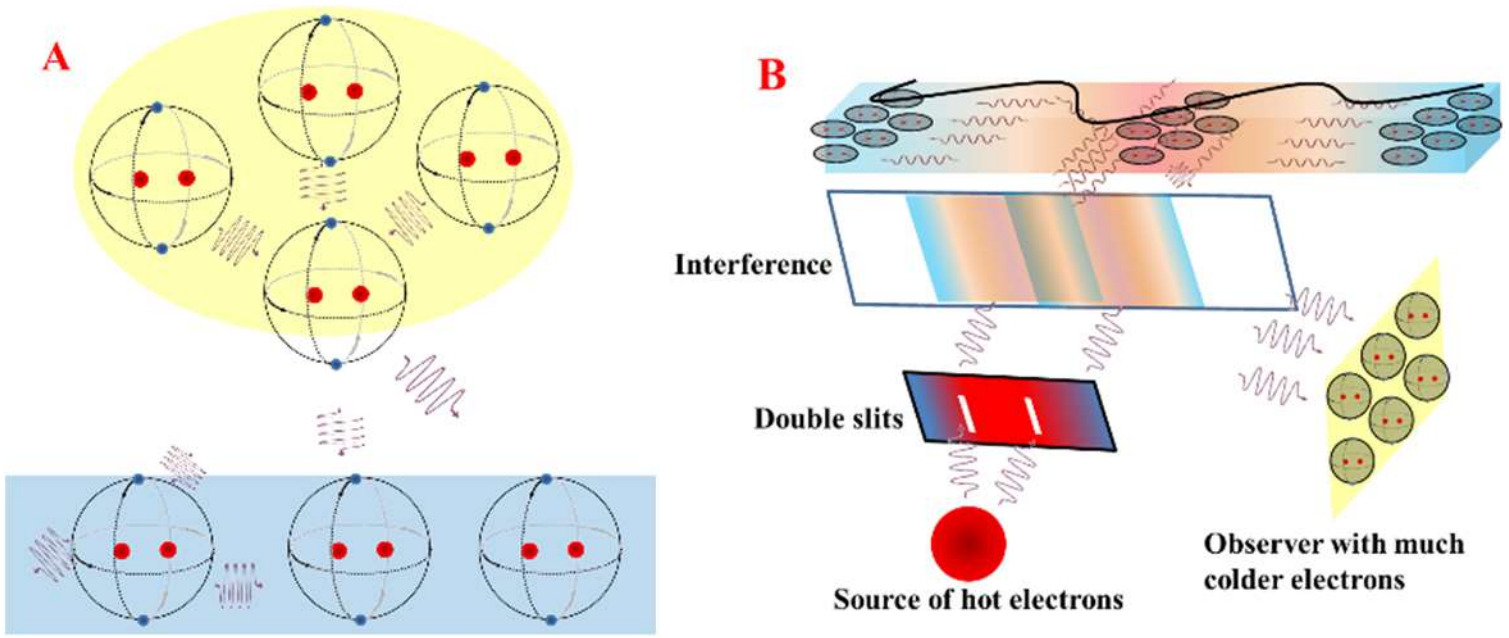

Figure 19, Light emission, adsorption, and light double slits interference

## Error of measurements

In conventional physics, the effect of the relative thermal difference has not been considered in the measurement of distance, velocity, and time. Here, we introduce the relative thermal difference into the measurements. As shown in Fig. 20, the observed object (A) is assumed to emit light, while a stationary observer (O) receives it. Meanwhile, B also emits light toward O, but its mass is negligible. When A shields the light from B, O infers that A has attained B. Directed by a compass, A walks with a constant $v_0$ along the preset direction of a straight line, AB. Due to the relative thermal difference between A and O, when O infers that A has attained B, the distance measured by O is AC.

$$AB = AC \cdot cos\frac{\theta}{2} \tag{70}$$

Otherwise, if A is directed by the light received from B, A must continuously adjust the direction of the compass while walking.

$$v_0 = v'cos\frac{\theta}{2} - \frac{AB}{2}tan\frac{\theta}{2} \cdot \frac{d\theta}{dt} \tag{71}$$

$v'$ is the measured velocity. If $AB = v_0 dt$,

$$\frac{v'}{v_0} = \frac{2+tan\frac{\theta}{2} \cdot d\theta}{2cos\frac{\theta}{2}} \tag{72}$$

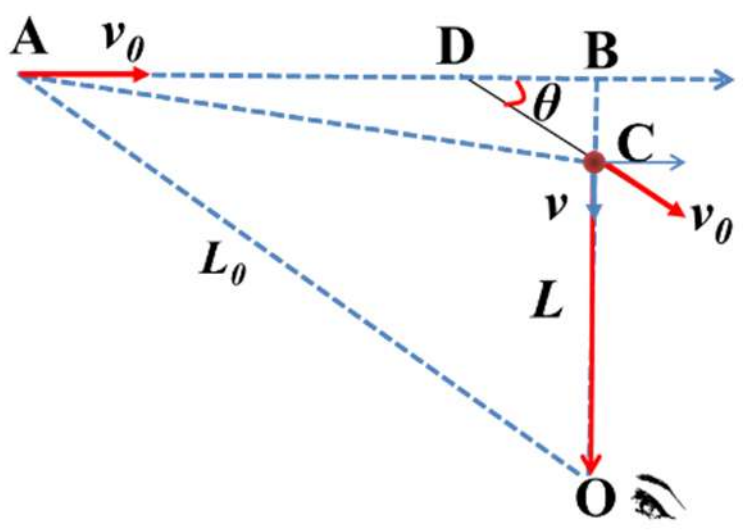

Figure 20, Relative thermal difference in the measurement of velocity

## Interchange of Temperatures: Molecular, Nuclear, and Electron Temperatures

In a gas system, the alignments of molecules are random. However, as in typical cases, Fig. 20

shows the bond-orbital alignments of two molecules, i.e., $P_1$ and $P_2$. Cases a) denote the parallel planes, b) the orbits on the mutually vertical planes, and c) the coplanar orbitals. Case a) contains two subcases, i.e., the equidirectional rotation and counter-rotation. Given $L>L_e$, $P_1$ and $P_2$ are mutually attractive in the equidirectional rotation. If $P_1$ is hotter with the shorter AC and higher frequency $f_A$, A and C particles accelerate B and D particles. As a result, AC=BD and $f_A$=$f_B$. However, as $L<L_e$, $P_1$ and $P_2$ are mutually repulsive; AC shortens while BD elongates. Meanwhile, $f_A$ rises while $f_B$ decays. In the counter-rotation, both AC and BD elongate. Meanwhile, $f_A$ and $f_B$ decay regardless of the distance. As a result, the bonds may break and recombine.

Case b) concerns the distance regardless of the rotational direction, as $L > L_e$, the hotter particles transfer heat to the colder ones. However, when $L < L_e$, the bonds may break and reform.

Case c) has only one result: the isothermal orbits turn into non-isothermal orbits, regardless of the distance and the rotational directions.

The orbits of bonded electrons change simultaneously with the change of bonded nuclei.

Since the attractive force is inversely proportional to $L$, the forces increase as the volume shrinks. As discussed above, attractive forces play a role in converting the kinetic and potential energies of bonded nuclei and electrons. The sum of potentials relates to α in Eq. 3. As the distance is short enough, the repulsive force also works for the conversion, but the result is opposite to that of the attractive force.

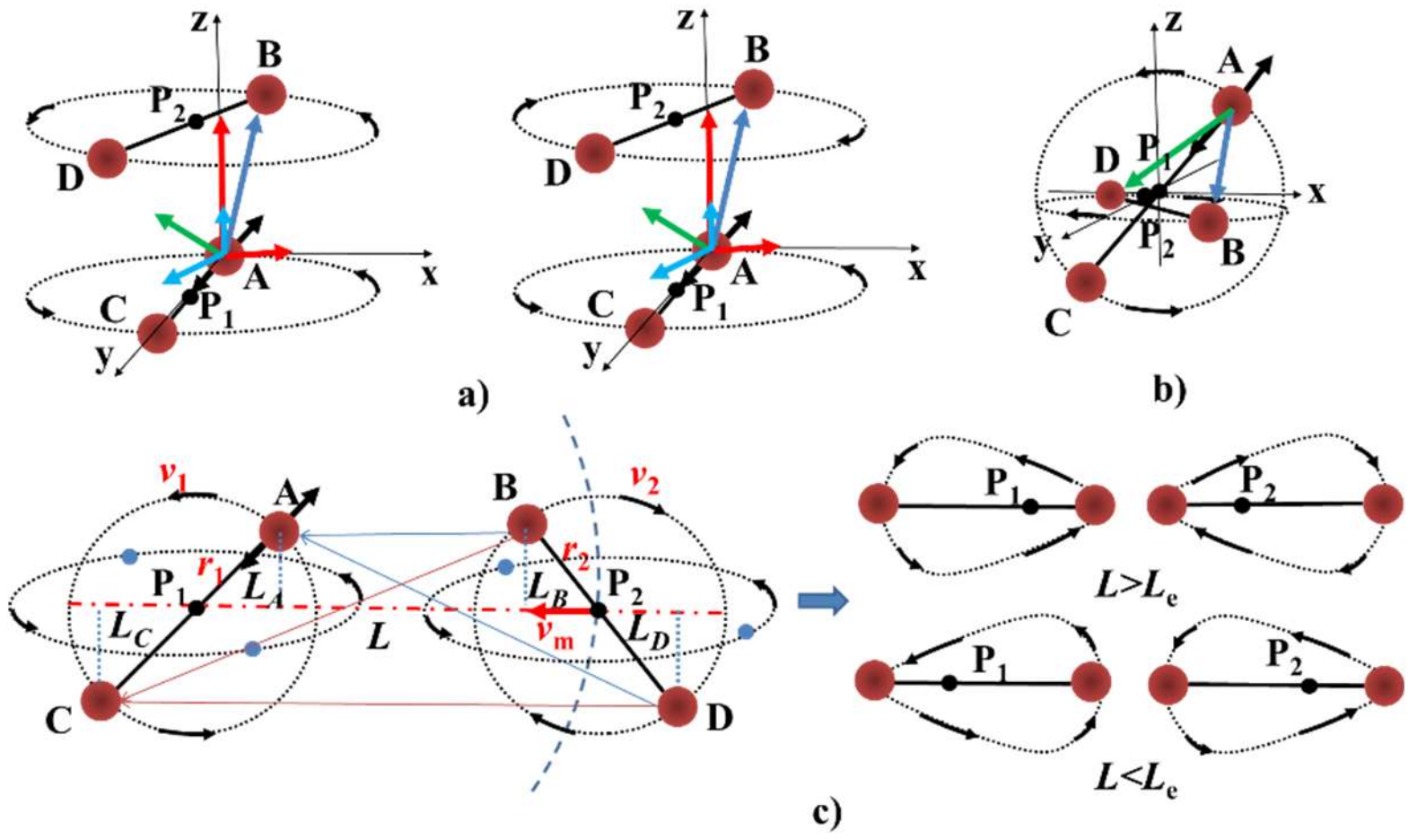

Figure 20, Heat transfer between two molecules

In a random system, minor cases are the norm. The majority are the deformed cases shown in Fig. 20, in which the BD plane crosses the AC plane at an angle. In Case a), the potential increases only in counter-rotation. In the equidirectional rotation, the potentials remain unchanged. In Case b), $L>L_e$, the potential decreases, whereas $L<L_e$, the potential increases. In Case c), the potentials of the two molecules are constant. Hence, as $L>L_e$, the probability is $\frac{1}{6}$ increase and $\frac{1}{6}$ decrease, whereas, as $L>L_e$, it is $\frac{1}{2}$, in which the potential increases. That is why the system's potential dramatically increases near $T_c$. [29]

Provided that the system is composed of a single type of molecule and the molecular temperature is constant, the repulsive force is Coulomb's force between two nuclei.

[Continue…]

## Conclusion

A new thermodynamic framework shapes modern physics: Electrons condense to form hot protons and cool protons (neutrons), then construct the nucleus through a thermal-difference force. An atom is an adiabatic electron-gas system. Heat is the sum of variations in kinetic energy, system potential, and environmental potential within a system. Force is the heat gradient over distance in the direction of transfer. Two kinds of forces exist between two particles, resulting from the relative thermal difference and the relative difference in number-volume density. Mutual repulsion and attraction simultaneously exist. Interaction doesn't need a medium. As a result, a series of concise equations is derived that includes only spatial parameters, mass, and temperature. Electricity and magnetism are nonessential concepts. 1) The equilibrium distance of molecular interaction: $L_e = \sqrt[3]{\frac{3\pi^{\alpha-1}m_A g}{4N_A kT}}$; 2) the atom structure is built based on the radii of atoms and the temperature of electrons. An atom has only one isothermal orbital and several non-isothermal orbitals constituting a tetrahedron. Each orbital accommodates 2 electrons with opposite spins; 3) the 1st ionization energies (IE) are close to the kinetic energies of surface electrons, $\frac{3}{2}kT_e$; 4) the strengths of chemical bonds and H-bonds are calculated, e.g., the strength of O-O: $\Delta E = 2 \cdot \frac{3k}{2}(\frac{2}{\sqrt{3}} - 1)T_0$; 5) Light is heat transferred from a hot electron to a cold electron. The rate of hot electrons determines the velocity, but the hot molecule's rotation rate determines the frequency. For example, a $H_2$ on Earth transfers heat to a cold particle with velocity, $v_{exo} = 9.59 \times 10^7 sin(\pi f_e t)\cos(\pi f_e t)\cos(2\pi f_p t)$. Hence, the energy is transferred by piece and piece. This is the so-called wave-particle duality.

## Acknowledgments

The paper was completed independently by the author without any supporting funds.

## Conflict of Interests

The authors declare no conflict of interest.

## References

[1]. Wikipedia: https://en. Wikipedia. org/wiki/Big-Bang
[2]. J. D. Van der Waals, *Over de Continuiteit van den Gasen Vloeistoftoestand*, Sijthoff, 1873; Vol. 1.
[3]. W. Sutherland, *London, Edinburgh Dublin Philos. Mag. J. Sci.* 1893, 36, 507−531. LII. The viscosity of gases and molecular force.
[4]. P. Drude, *Ann. Phys.* 1900, 306, 566−613. Zur Elektronentheorie der Metalle.
[5]. W. L. Bade, *J. Chem. Phys.* 1957, 27, 1280−1284. Drude-Model Calculation of Dispersion Forces. I. General Theory.
[6]. J. O. Hirschfelder, C. F. Curtiss, R. B. Bird, *The Molecular Theory of Gases and Liquids*; John Wiley & Sons: New York, 1964.

[7]. F. London, *Z. Phys.* 1930, 63, 245−279. Zur Theorie und Systematik der Molekularkrafte.
[8]. F. London, *Trans. Faraday Soc.* 1937, 33, 8b−26. The general theory of molecular forces.
[9]. G. Mie, *Ann. Phys.* 1903, 316, 657−697. Zur kinetischen Theorie der einatomigen Körper.
[10]. E. Grüneisen, *Ann. Phys.* 1912, 344, 257−306. Theorie des festen Zustandes einatomiger Elemente.
[11]. J. E. Jones, *Proc. R. Soc. London, Ser. A* 1924, 106, 441−462. On the Determination of Molecular Fields. I. From the Variation of the Viscosity of a Gas with Temperature.
[12]. J. E. Jones, *Proc. R. Soc. London, Ser. A* 1924, 106, 463−477. On the Determination of Molecular Fields. II. From the Equation of State of a Gas.
[13]. J. E. Jones, *Proc. R. Soc. London, Ser. A* 1924, 106, 709−718. On the Determination of Molecular Fields. III. From crystal measurements and kinetic theory data.
[14]. P. Schwerdtfeger, D.J. Wales, *J. Chem. Theory Comput.* 2024, 20, 3379−3405. 100 Years of the Lennard-Jones Potential.
[15]. M. P. Allen, D. J. Tildesley, *Computer Simulation of Liquids*; Clarendon Press: New York, 1989.
[16]. D. Frenkel, B. Smit, *Understanding Molecular Simulation: From Algorithms to Applications*, 2nd ed.; Academic Press: London, 2002.
[17]. S. Deffner, S. Campbell, *Quantum Thermodynamics: An introduction to the thermodynamics of quantum information*, Morgan & Claypool Publishers, 2019.
[18]. F. Binder, L.A. Correa, C. Gogolin, J. Anders, G. Adesso, *Thermodynamics in the Quantum Regime. Fundamental Theories of Physics,* Springer, 2018.
[19]. B. Poirier, *A Conceptual Guide to Thermodynamics,* Wiley, 2014.
[20]. J. Güémez, J. A. Mier, *Eur. J. Phys.* 2024, 45, 015701. Relativistic mechanics and thermodynamics: IV. Thermodynamic processes.
[21]. O. Redlich, J. N. S. Kwong, *Chem. Rev.* 1949 44 (1): 233–244. On the Thermodynamics of Solutions. V. An Equation of State. Fugacities of Gaseous Solutions.
[22]. D. Y. Peng, D. B. Robinson, *Ind. Eng. Chem.: Fundamentals*. 1976 15: 59–64. A New Two-Constant Equation of State.
[23]. R. Stryjek, J. H. Vera, *Can. J. Chem. Eng.* 1986, 64 (5): 820–826. PRSV2: A cubic equation of state for accurate vapor-liquid equilibria calculations.
[24]. A. Kaplun, A. Meshalkin, *EPJ Web of Conferences*, 2014, 76, 01026. Unified low-parametrical equation of state for engineering calculations of thermodynamic properties of substances.
[25]. J. D. Dymond, R. C. Wilhoit, *Virial coefficients of pure gases and mixtures*, Springer, 2003.
[26]. M. Benedict, G.B. Webb, L. C. Rubin, *J. Chem. Phys.*, 1940, 8 (4): 334–345. An Empirical Equation for Thermodynamic Properties of Light Hydrocarbons and Their Mixtures: I. Methane, Ethane, Propane, and n-Butane.
[27]. Wikipedia: https://en.wikipedia.org/wiki/Newton%27s_law_of_universal_gravitation.
[28]. Van W. Gordon, S. Richard, *Fundamentals of Classical Thermodynamics (2nd edition). Chapter 4.7, Definition of Heat*, John Wiley & Sons, 1978.
[29]. H.-M. Ni, B. Zhu, *ChemRxiv*, 2022, *Preprint.* DOI: 10.26434/chemrxiv-2022-5m718-v4. A general and simple state equation for all gases.
[30]. H.-M. Ni, *J. Mol. Liquids*, 2022, 368, 120690. On the Hydrophobic Hydration, Solvation and Interface: A Thought Essay (I).

[31]. X.-Y. Cui, B. Zhu, H.-M. Ni, *ChemistrySelect*, 2024, 9, e202304699 (1 of 10). A New Model for Specific Heat Capacity of Real Gas.
[32]. H.-M. Ni, *ChemRxiv*, 2025, *Preprint*. 10.26434/chemrxiv-2025-dtd2t. On Molecule Symmetry, Latent Heat, and Entropy.
[33]. W. G. Hoover, C. G. Hoover, K. P. Travis, *Phys. Rev. Lett.* 2014, 112 (14): 144504. Shock-Wave Compression and Joule–Thomson Expansion.
[34]. S. S. Batsanov, *Inorg. Mat.* 2001, 37 (9): 871–885. Van der Waals Radii of Elements.
[35]. J. Fischer, M. Wendland, *Fluid Phase Equilibria*, 2023, 573, 113876. On the history of key empirical intermolecular potentials.
[36]. J. Fischer, R. Lustig, H. Breitenfelder-Manske, W. Lemming, *Mol. Phys.* 1984, 52(2), 485-497. Influence of intermolecular potential parameters on orthobaric properties of fluids consisting of spherical and linear molecules.
[37]. Y. Song, E. A. Mason, *Phys. Rev. A*, 1990, 42(8), 4749-4755. Analytical equation of state for molecular fluids: Comparison with experimental data.
[38]. R. Lustig, *Ber. Bunsenges. Phys. Chem.* 1994, 98 (5), 706- 711. Thermodynamics of Liquid Oxygen from Molecular Dynamics Simulations.
[39]. T. Boulik, *J. Chem. Phys.* 1987, 87, 1751-1756. Simple perturbation method for convex-molecule fluids
[40]. N. Yu, A. A. Polycarpou, *J. Coll. Interf. Sci.* 2004, 278, 428-435. Adhesive contact based on the Lennard–Jones potential: a correction to the value of the equilibrium distance as used in the potential.
[41]. G. Lamanna, J. van Poppel, M.E.H. van Dongen, *Experiments in Fluids*, 2002, 32, 381-395. Experimental Determination of Droplet Size and Density Field in Condensing Flows.
[42]. A. W. Adamson, A. P. Gast., *Physical chemistry of surfaces*; 6 Ed, Wiley, 1997.
[43]. G. P. Bewley, D. P. Lathrop, K. R. Sreenivasan, *Nature*, 2006, 441, 588. Superfluid Helium: Visualization of quantized vortices.
[44]. A. Kheradvar, G. Pedrizzetti, *Vortex Dynamics. In: Vortex Formation in the Cardiovascular System.* Springer, London, 2012. https://doi.org/10.1007/978-1-4471-2288-3_2.
[45]. E. Clementi, D.L. Raimondi, W.P. Reinhardt, *J. Chem. Phys.* 1967,47 (4): 1300–1307. Atomic Screening Constants from SCF Functions. II. Atoms with 37 to 86 Electrons.
[46]. NIST Chemistry WebBook, https://webbook.nist.gov/chemistry/.
[47]. B. deB. Darwent, *Nat. Stand. Ref. Ser., Nat. Bur. Stand.* (USA), NSRDS-NBS 31, 1970. Bond Dissociation Energy in Simple Molecules.
[48]. Computational Chemistry Comparison and Benchmark Data Base, NIST, https://cccbdb.nist.gov/diatomicexpbondx.asp. List of Experimental Diatomic Bond Lengths.
[49]. J. J. Krauth, K. Schuhmann, M. A. Ahmed, F. D. Amaro, P. Amaro, F. Biraben, T.-L. Chen, D. S. Covita, A. J. Dax, M. Diepold, L. M. P. Fernandes, B. Franke, S. Galtier, A. L. Gouvea, J. Götzfried, T. Graf, T. W. Hänsch, J. Hartmann, M. Hildebrandt, P. Indelicato, L. Julien, K. Kirch, A. Knecht, Y.-W. Liu, J. Machado, C. M. B. Monteiro, F. Mulhauser, B. Naar, T. Nebel, F. Nez, J. M. F. dos Santos, J. P. Santos, C. I. Szabo, D. Taqqu, J. F. C. A. Veloso, J. Vogelsang, A. Voss, B. Weichelt, R. Pohl, A. Antognini, F. Kottmann, *Nat.* 2021, 589, 527-531. Measuring the α-particle charge radius with muonic helium-4 ions.